\documentclass{aa}
\usepackage[utf8]{inputenc}
\usepackage{natbib}
\usepackage[flushleft]{threeparttable}
\usepackage{pdflscape}
\usepackage{tikz}
\usepackage{graphicx}
\usepackage{hyperref}
\usepackage{xcolor}
\usepackage{float}
\usetikzlibrary{positioning}
\bibpunct{(}{)}{;}{a}{}{,} % to follow the A&A style

\usepackage{amstext}

\begin{document}

\title{X-Shooting ULLYSES: Massive stars at low metallicity }
\subtitle{III. Terminal wind speeds of ULLYSES massive stars}

\titlerunning{X-Shooting ULLYSES -- III. Terminal wind speeds of ULLYSES massive stars}

\author{C. Hawcroft\inst{\ref{inst1},\ref{inst2}}\and H. Sana\inst{\ref{inst1}} \and L. Mahy\inst{\ref{inst3}} \and J.O. Sundqvist\inst{\ref{inst1}} \and A. de Koter\inst{\ref{inst4},\ref{inst1}} \and P.A. Crowther\inst{\ref{inst5}} \and J.M. Bestenlehner\inst{\ref{inst5}} \and S.A. Brands\inst{\ref{inst4}} \and A. David-Uraz\inst{\ref{inst6},\ref{inst7}} \and L. Decin\inst{\ref{inst1}} \and C. Erba\inst{\ref{inst8}} \and M. Garcia\inst{\ref{inst9}} \and W.-R. Hamann\inst{\ref{inst10}} \and A. Herrero\inst{\ref{inst11},\ref{inst12}} \and R. Ignace\inst{\ref{inst8}} \and N. D. Kee\inst{\ref{inst13}} \and B. Kub\'{a}tov\'{a}\inst{\ref{inst14}} \and R. Lefever\inst{\ref{inst15}} \and A. Moffat\inst{\ref{inst16}} \and F. Najarro\inst{\ref{inst9}} \and L. Oskinova\inst{\ref{inst10}} \and D. Pauli\inst{\ref{inst10}} \and R. Prinja\inst{\ref{inst17}} \and J. Puls\inst{\ref{inst18}} \and A.A.C. Sander\inst{\ref{inst15}} \and T. Shenar\inst{\ref{inst4}} \and N. St-Louis\inst{\ref{inst16}} \and A. ud-Doula\inst{\ref{inst19}} \and J.S. Vink\inst{\ref{inst20}}}
\institute{Institute of Astronomy, KU Leuven, Celestijnenlaan 200D, 3001, Leuven, Belgium \label{inst1} \and Space Telescope Science Institute, 3700 San Martin Drive, Baltimore, MD 21218, USA \\ email: chawcroft@stsci.edu \label{inst2} \and Royal Observatory of Belgium, Avenue Circulaire 3, B-1180 Brussels, Belgium \label{inst3} \and Astronomical Institute Anton Pannekoek, Amsterdam University, Science Park 904, 1098 XH Amsterdam, The Netherlands \label{inst4} \and Department of Physics and Astronomy, University of Sheffield, Hicks Building, Hounsfield Road, Sheffield, S3 7RH, UK \label{inst5} \and Department of Physics and Astronomy, Howard University, Washington, DC 20059, USA\label{inst6} \and Center for Research and Exploration in Space Science and Technology, and X-ray Astrophysics Laboratory, NASA/GSFC, Greenbelt, MD 20771, USA \label{inst7} \and Department of Physics \& Astronomy, East Tennessee State University, Johnson City, TN 37614, USA \label{inst8} \and Centro de Astrobiolog\'{i}a, CSIC-INTA. Crtra. de Torrej\'{o}n a Ajalvir km 4, E-28850 Torrej\'{o}n de Ardoz (Madrid), Spain \label{inst9} \and Institut f{\"u}r Physik und Astronomie, Universit{\"a}t Potsdam, Karl-Liebknecht-Str. 24/25, 14476 Potsdam, Germany \label{inst10} \and Instituto de Astrof\'{i}sica de Canarias, C/ V\'{i}a L\'{a}ctea s/n, E-38200 La Laguna, Tenerife, Spain \label{inst11} \and Departamento de Astrofisica, Universidad de La Laguna, 38205, La Laguna, Tenerife, Spain \label{inst12} \and National Solar Observatory, 22 Ohi`a Ku St, Makawao, HI 96768, USA \label{inst13} \and Astronomick\'{y} \'{u}stav, Akademie v\v{e}d \v{C}esk\'{e} republiky, CZ-251 65 Ond\v{r}ejov, Czech Republic \label{inst14} \and Zentrum f{\"u}r Astronomie der Universit{\"a}t Heidelberg, Astronomisches Rechen-Institut, M{\"o}nchhofstr. 12-14, 69120 Heidelberg, Germany \label{inst15} \and D\'{e}partement de Physique and Centre de Recherche en Astrophysique du Qu\'{e}bec (CRAQ) Universit\'{e} de Montr\'{e}al, C.P. 6128, Succ. Centre-Ville, Montr\'{e}al, Qu\'{e}bec, H3C 3J7, Canada \label{inst16} \and Department of Physics and Astronomy, University College London, Gower Street, London WC1E 6BT, UK \label{inst17} \and LMU M{\"u}nchen, Universit{\"a}tssternwarte, Scheinerstr. 1, 81679 M{\"u}nchen, Germany \label{inst18} \and Dept. of Physics, Penn State Scranton, 120 Ridge View Drive, Dunmore, PA 18512, USA \label{inst19} \and Armagh Observatory and Planetarium, College Hill, BT61 9DG Armagh, Northern Ireland \label{inst20}}

\abstract{
    % context (optional)
The winds of massive stars have a significant impact on stellar evolution and on the surrounding medium. The maximum speed reached by these outflows, the terminal wind speed $v_{\infty}$, is a global wind parameter and an essential input for models of stellar atmospheres and feedback. With the arrival of the ULLYSES programme, a legacy UV spectroscopic survey with the Hubble Space Telescope, we have the opportunity to quantify the wind speeds of massive stars at sub-solar metallicity (in the Large and Small Magellanic Clouds, $0.5Z_{\odot}$ and $0.2Z_{\odot,}$ respectively) at an unprecedented scale.
}{
    % aims
We empirically quantify the wind speeds of a large sample of OB stars, including supergiants, giants, and dwarfs at sub-solar metallicity. Using these measurements, we investigate trends of $v_{\infty}$ with a number of fundamental stellar parameters, namely effective temperature ($T_{\rm{eff}}$), metallicity ($Z$), and surface escape velocity $v_\mathrm{esc}$. 
}{
    % methods
We empirically determined $v_{\infty}$ for a sample of 149 OB stars in the Magellanic Clouds either by directly measuring the maximum velocity shift of the absorption component of the \ion{C}{IV} $\lambda\lambda$1548-1550 line profile, or by fitting synthetic spectra produced using the Sobolev with exact integration method. Stellar parameters were either collected from the literature, obtained using spectral-type calibrations, or predicted from evolutionary models.
}{
    % results
We find strong trends of $v_{\infty}$ with $T_{\rm{eff}}$ and $v_\mathrm{esc}$ when the wind is strong enough to cause a saturated  P Cygni profile in \ion{C}{IV} $\lambda\lambda$1548-1550. We find evidence for a metallicity dependence on the terminal wind speed $v_{\infty} \propto Z^{0.22\pm0.03}$ when we compared our results to previous Galactic studies.
}{ Our results suggest that $T_{\rm{eff}}$ rather than $v_\mathrm{esc}$ should be used as a straightforward empirical prediction of $v_{\infty}$ and that the observed $Z$ dependence is steeper than suggested by earlier works.
}

\maketitle

\section{Introduction} \label{sec: introduction}

Hot, massive stars ($> 8 M_{\odot}$) are known to host radiatively driven outflows. The strong ultraviolet (UV) flux of these stars transfers enough momentum outward through scattering with or absorption by UV metal ion line transitions to remove material from the stellar surface \citep{Lucy1970, Castor1975}. These winds are an essential factor in the evolution of the star as they can reduce the stellar mass considerably throughout the stellar lifetime, thereby altering both the evolutionary pathway \citep{Ekstrom2012} and the properties of the end products \citep{Langer2012, Smith2014}.

Typically, two fundamental global parameters are used to describe the winds of massive stars: the mass-loss rate ($\dot{M}$), and the asymptotic or terminal speed of the outflow ($v_{\infty}$) \citep{Puls2008, Hillier2020, Vink2021}. When the winds are adiabatic, the wind luminosity ($0.5\dot{M}v_{\infty}^{2}$) determines the efficiency of the wind feedback\footnote{If the majority of energy from winds is lost via radiative cooling, the direct injection of momentum from winds becomes the governing factor ($\dot{M}v_{\infty}$), which is considerably weaker than the ($0.5\dot{M}v_{\infty}^{2}$) factor for adiabatic winds \citep{Silich2013}.} \citep{Weaver1977}. Ionising radiation from stars \citep{Spitzer1978, Dale2012} also plays an important role that often combines in non-linear ways with wind feedback \citep{Geen2022}. 

Resonance transitions at UV wavelengths of species such as carbon are prime diagnostics of the terminal velocity, especially when the spectral lines remain optically thick at the distances where the wind velocity reaches $v_{\infty}$. The advantage of this is that the terminal wind speed can be measured directly from UV spectroscopy. This measurement is independent of all other stellar parameters because we observe an extended region of zero flux in the P Cygni profile of the resonance line, the farthest extent of which is associated with the maximum velocity of wind material available to block the emergent light \citep{Prinja1990}. We also observe a rapid increase from zero to continuum flux levels at the blue edge of the resonance-line profile. The gradient of the profile edge can be reduced through high-velocity turbulent gas generated by shocks coming from collisions between high-velocity low-density material and low-velocity high-density material and can be used to diagnose the characteristic velocity dispersion of the outflow. The formation of this multi-component wind is inherent to massive stars through the instability associated with radiation line-driving \citep{MacGregor1979, Carlberg1980, Owocki1984, Feldmeier1995, Driessen2019}.

Absorption troughs of resonance-line profiles may not reach zero flux for all stars. For stars without P Cygni saturation, a wind speed can still be measured from the blue edge of the line profile, but it will then not reflect the terminal wind speed, but rather the maximum observed wind speed at the highest velocity of the absorption profile, or a lower limit to $v_{\infty}$. The wind speeds can also be measured by fitting simplified models, which implement the Sobolev with exact integration method (SEI) from \cite{Lamers1987}, and are significantly less computationally expensive than full stellar atmosphere and wind models. 
There may be some offset between wind speeds that are directly measured from the trough and those obtained through fitting. By applying the SEI fitting method to all stars in the sample, we are able to quantify possible discrepancies between SEI and DM values and comment on the uncertainty of the terminal wind speeds from unsaturated resonance lines.

Line-driven wind theory predicts that the strength of stellar winds changes with metallicity \citep{Puls2000}, so there is a need to empirically quantify the wind strength in environments with metallicities different than that of the Milky Way. Theoretical \citep{Vink2001, Krticka2018, Bjorklund2021} and empirical \citep{Mokiem2007, Ramachandran2019} studies have been conducted to quantify the $\dot{M}(Z)$ dependence of OB stars. Empirical studies have shown that it is not straightforward to quantify the $v_{\infty}(Z)$ dependence with observations of massive stars that are currently available because while $v_{\infty}$ has been determined for 250 OB stars \citep{Prinja1990}, the parameter space coverage of stars at low metallicity with prominent, observable UV wind line profiles is limited \citep{Haser1995thesis, Garcia2014, Marcolino2022}. As a result of this, the only predictions of $v_{\infty}(Z)$ are theoretical \citep{Leitherer1992, Krticka2018, Bjorklund2021, VinkSander2021}, with the most widely used prediction being $\sim{} Z^{0.13}$ from \cite{Leitherer1992}. The Hubble Space Telescope (HST) UV Legacy Library of Young Stars as Essential Standards (ULLYSES; \citealp{RomanDuval2020}) programme changes this situation, providing UV spectroscopy of a large sample of O, early-B, mid- to late-B supergiant and Wolf-Rayet type stars spanning the upper Hertzsprung-Russell diagram (HRD) in the Large and Small Magellanic Clouds (LMC and SMC). These two satellite galaxies represent environments with metallicities that are substantially lower than that of the Milky Way; the metallicity of the LMC is one-half and that of the SMC is one-fifth of the solar metallicity \citep{Mokiem2007}.

The opportunity arises with this dataset (if sufficient characterisation of the stars is available) to explore trends in the wind speed with fundamental stellar parameters, which might allow for future empirical predictions of the terminal wind speed given other fundamental stellar parameters such as the luminosity and effective temperature. This has been done for Galactic stars and is discussed throughout Sect. \ref{sec: discussion} (see e.g. \citealp{Kudritzki2000} and references therein). This could help to make a well-educated assumption for $v_{\infty}$ in stellar atmosphere studies that lack a $v_{\infty}$ diagnostic \citep{Bouret2015}. Conversely, these trends could be used to estimate stellar parameters from terminal wind speeds, allowing for calibrations of stellar parameters without extensive atmosphere modelling. This may only be useful when only $v_{\infty}$ is known, however, because a spectral type classification will likely give a better estimation of the stellar parameters. Here we present an analysis of the terminal wind speeds of a sample of 149 OB stars in the LMC and SMC by investigating trends with fundamental stellar parameters including effective temperature, surface escape speed, and metallicity. 

The paper is laid out as follows: Section \ref{sec: sample} introduces the ULLYSES sample and the associated UV observations obtained with the HST. Section \ref{sec: methods} outlines our method for measuring wind speeds and obtaining or estimating other stellar parameters. In Section \ref{sec: results} we present the results of this analysis, including the terminal wind speeds. In Section \ref{sec: discussion} we discuss the relations between terminal wind speeds and other stellar parameters, and we compare them to similar studies in the Galaxy. Our conclusions are given in Section \ref{sec: conclusions}.

\section{Sample and observations} \label{sec: sample}

\begin{figure*}[t!]
    \centering
    \includegraphics[scale=0.58]{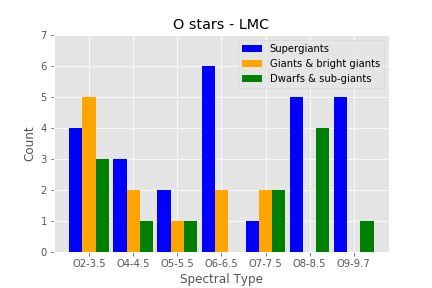}
    \includegraphics[scale=0.58]{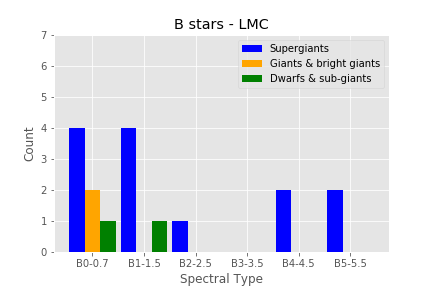}
    \includegraphics[scale=0.58]{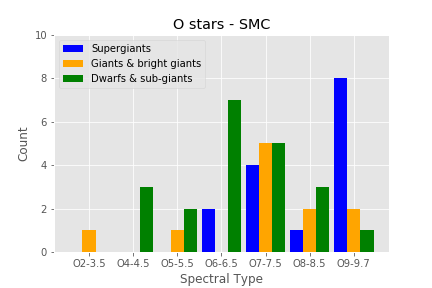}
    \includegraphics[scale=0.58]{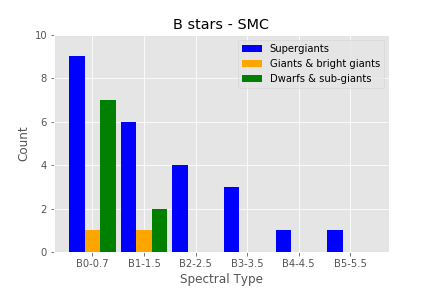}
    \caption{Histogram of the distribution of the stellar sample through spectral type and luminosity class. LMC stars are shown in the first and second panels, which correspond to the O- and B-type stars, respectively. Similarly, SMC stars are presented in the third and fourth panels.}
    \label{fig: histogram}
\end{figure*}

The sample used in this work was obtained as part of the ULLYSES programme \citep{RomanDuval2020}, which aims to secure high-quality UV spectra of massive stars in the LMC and SMC, sampling all spectral sub-types and luminosity classes from O2 to B1.5, as well as supergiants in the range of B2 to B9 with redundancy. This study uses spectra up to the third ULLYSES data release. This includes 67 stars in the LMC and 82 stars in the SMC. The LMC sample  comprises 39 OB supergiants, 14 OB giants, and 14 OB dwarfs. The SMC sample includes 39 OB supergiants, 13 OB giants, and 30 OB dwarfs. The distribution of targets by spectral type and luminosity class is shown in Fig. \ref{fig: histogram}. The full list of objects is presented in Tables \ref{tbl:Results-LMC-sgiants} to \ref{tbl:Results-SMC-dwarfs}. 

\subsection{UV data} \label{sec: uv-data}

The data were obtained with either the Cosmic Origins Spectrograph (COS) or with the Space Telescope Imaging Spectrograph (STIS), using the appropriate grating required to obtain sufficient coverage of relevant diagnostics, depending on the target spectral type. The target signal-to-noise ratio (S/N) differs slightly for each grating; generally, \footnote{The target S/N is dependent on a combination of brightness, wavelength, and spectral type. The full list of criteria can be found on the ULLYSES website:  \href{https://ullyses.stsci.edu/index.html\#generalinformation}{\textcolor{blue}{https://ullyses.stsci.edu/index.html\#generalinformation}}} S/N $\approx$ 20-30. This essentially means that the bright targets are observed with STIS and the fainter stars are observed with COS, which offers better sensitivity at the cost of spectral resolution. The COS gratings cover the spectral range from 937 to 1792 \AA\ for O2-O9 stars and 1830 to 2110 \AA\ for B2-B9 stars; the STIS gratings run from 1141 to 2366 \AA\ for O2-B9 stars and from 2275 to 3119 \AA\ for B5-B9 stars (for which only supergiants were observed).

\subsection{Stellar parameters} \label{sec: bonnsai}

We adopted spectral types, luminosities, and effective temperatures from the ULLYSES metadata, comprised of both literature values and spectral type calibrations (\cite{RomanDuval2020}; full metadata are available on the \href{https://ullyses.stsci.edu/ullyses-data-description.html}{\textcolor{blue}{ULLYSES website}}, latest access date: 5 April 2022). In order to estimate other stellar parameters, we compared luminosity and effective temperature to the single-star evolution models of \cite{Brott2011} and \cite{Kohler2015} with the Bayesian analysis tool BONNSAI \citep{Schneider2014}. This provides us with evolutionary masses and radii that allow us to compute the initial and current surface escape speeds, 
\begin{equation} \label{vesc-eqn}
v_{\rm{esc}} = (2\mathrm{G}M(1-\Gamma_{\rm{e}})/R)^{1/2},
\end{equation}
where G is the gravitation constant, $M$ is the stellar mass, $R$ is the stellar radius, and $\Gamma_{\rm{e}}$ is the Eddington parameter: the ratio of radiative acceleration to local gravitational acceleration, here assuming the radiative component to be purely electron scattering. Throughout this paper, any mention of surface escape speed refers to surface escape speeds reduced by electron scattering, as defined here. We highlight there will be additional uncertainties on the escape speeds used here as stellar evolutionary tracks are shaped by a number of other stellar parameters, such as the rotational velocity and mass-loss rate, which will affect the BONNSAI results. For example, a reduction in mass-loss rate on the main sequence by a factor of 3 or more would introduce a systematic correctional shift to higher evolutionary masses. Moreover, stellar properties such as the mass and radius will vary when different stellar evolutionary codes are used (see e.g. \citealp{Agrawal2021}).

\section{Methods} \label{sec: methods}

We employed two methods to measure the terminal wind speeds. The initial method, direct measurement (DM), can be used for stars with saturated resonance-line profiles \citep{Prinja1990}. Direct measurement can also be applied to stars with unsaturated profiles, but then only a lower limit on $v_{\infty}$ can be obtained. The second method, SEI modelling, is needed for stars without saturation in their UV resonance-line profiles, although it can also be used for stars with saturated profiles \citep{Lamers1987}. SEI models are further needed to constrain the turbulent velocity of the wind ($v_{\rm{turb}}$). For either method, minimal data processing is required from the ULLYSES high-level science product spectra; only a local normalisation around the spectral line of interest was carried out.

A few caveats must be pointed out. We did not consider the radial velocity of individual stars when fitting the profiles because the lack of isolated absorption line profiles in the UV makes these measurements difficult. We only compensated for the radial velocity associated with the host Magellanic Cloud by applying a Doppler shift to our SEI model before comparing to the observed spectrum, 262 $\rm{km}\, \rm{s^{-1}}$ in the LMC, and 146 $\rm{km}\, \rm{s^{-1}}$ in SMC \citep{McConnachie2012}. An optical follow-up survey with the X-Shooter spectrograph on the European Southern Observatory (ESO) Very Large Telescope (VLT), named X-Shooting ULLYSES (\href{https://massivestars.org/xshootu/}{\textcolor{blue}{XShootU}}), will allow for radial velocity measurements that can be applied to the wind speeds afterwards (Vink et al. in prep.). In the meantime, we consider the impact of differences in radial velocity on our measurements of $v_{\infty}$ using preliminary radial velocity measurements for individual stars from the XShootU collaboration. Any necessary corrections due to stellar radial velocities are small because the largest deviations from the systemic velocity of the Cloud correspond to a shift of 100 $\rm{km}\, \rm{s^{-1}}$ in $v_{\infty}$. For the majority of the sample, the corrections are much smaller. Generally, any discrepancies in $v_{\infty}$ due to radial velocity are covered by our error estimates. There are also assumptions made in our SEI model, discussed in Section \ref{sec: sei}, that would cause issues if we attempted to fit the full P Cygni profile. These assumptions have little impact on our results here because we only examined a sub-section of the profile centred around the wavelength of the high-velocity edge of the absorption profile. 

We did not check the spectra for signatures of binarity. While undetected companions are possible, they are likely to be of relatively low luminosity and so would only provide weak signatures in the spectra. Companions like this would have a negligible impact on the strong P Cygni wind profiles used in this study.

Several strong resonance profiles are located in the covered wavelength ranges: N\,{\sc v}\,$\lambda \lambda$1239-1243, C\,{\sc iv}\,$\lambda \lambda$1548-1551, and Si\,{\sc iv}\,$\lambda \lambda$1394-1403 in the spectra of O- and early B-type stars, with C\,{\sc ii}\,$\lambda \lambda$1335-1336 and Al\,{\sc iii}\,$\lambda \lambda$1855-1863 appearing in the spectra of B0.7-B2 Ia stars. Unfortunately, the N\,{\sc v} line is strongly blended by interstellar Ly{$\alpha$} absorption. The separation of the doublet lines in Si\,{\sc iv} is 9\,\AA\ or 1930 km\,s$^{-1}$. This complicates a reliable determination of the terminal flow speed from the methods applied in this study. As the C\,{\sc ii} and Al\,{\sc iii} resonance lines appear in only a few stars, we limited our diagnostic in the present work to the strong C\,{\sc iv} doublet. The C\,{\sc iv}\,$\lambda \lambda$1548-1551 doublet used in this study is commonly saturated and is therefore a good and consistent diagnostic throughout the ULLYSES sample, but it is also possible to determine terminal wind speeds through other methods and diagnostics, which can be focused on in future studies. For example, a significant number of stars are complemented by archival observations made with the Far Ultraviolet Spectroscopic Explorer (FUSE), including additional diagnostics such as \ion{P}{v} 1118-1128 and \ion{C}{iii} 977, which may also show saturated or strong P Cygni profiles. Additionally, narrow absorption components (NACs) have been shown to be a good indicator of the terminal wind speed (e.g. \citealp{Prinja1990}), and a number of well-defined NACs can be seen throughout the ULLYSES spectra. 

\subsection{Direct measurement} \label{sec: dm}

The DM method is facilitated by the fact that for line profiles with extended saturation, the terminal wind speed can be measured directly. We did this by measuring the wavelength of the bluest edge of the P Cygni absorption trough showing zero flux, as established in \cite{Prinja1990}. 
Given the S/N of the observed spectra, discussed in Sect. \ref{sec: sample}, it can be illustrative to measure the wavelength of minimum absorption within the boundaries of flux variation allowed by the S/N. This velocity of minimum absorption deviates from the directly measured terminal wind speed by 60 km\,s$^{-1}$ on average, but can be as high as 200 km\,s$^{-1}$. Examples of the velocity of minimum absorption can be seen in the lower panels of Fig. \ref{fig: Profile-Green-LMC}, which show unsaturated line profiles. The difference between these two measurements is provided as the error on a directly measured terminal wind speed for a saturated profile. We took 100 km\,s$^{-1}$ as a conservative estimate on the uncertainty of the lower limit of the terminal wind speed found from non-saturated line profiles using the DM method.  

\subsection{SEI modelling} \label{sec: sei}

If the C\,{\sc iv}\,$\lambda \lambda$1548-1551 doublet is not saturated, the method used to constrain the wind speed is to reproduce this profile with synthetic spectra. This modelling can be made at relatively low computational expense, using the Sobolev with Exact Integration (SEI) method first developed by \cite{Lamers1987}. The SEI code applied in this work builds on the code of \cite{Haser1995} and is presented in \cite{Sundqvist2014}. In short, the method combines a computation of the source function with the \cite{Sobolev1960} escape probability method with a formal solution of the radiative transfer equation. The radial distribution of the opacity (or number density) of the ionisation species at hand is parametrised, including chaotic small-scale motions in the flow that cause clumping and porosity effects in both physical and velocity space. We did not use these velocity-porosity modifications because we are only interested in constraining $v_{\infty}$ and $v_{\rm{turb}}$. As we do not aim to reproduce the full morphology of the profile, we limited our input free parameters. We fixed the $\beta$ wind acceleration parameter to unity. This choice impacts the profile morphology close to the line centre, but this region is not assessed in our fit evaluation. We varied only four parameters: the terminal wind speed, the microturbulent velocity $v_{\rm{turb}}$ , and two parameters ($\kappa{}_{0}$ and $\alpha$) describing the line opacity. The latter is formalised as

\begin{equation}
\kappa(v) = \kappa_{0} \left(\frac{v}{v_\infty}\right)^{\alpha},
\end{equation}

\noindent where $\kappa{}_{0}$ is a dimensionless opacity parameter \citep{Hamann1981}. $\kappa{}_{0}$ and $\alpha$ were varied to match the strength of the absorption, therefore their magnitudes are not necessarily physically motivated. However, the turbulent wind speed can be constrained with the gradient of the return to continuum from the high-velocity edge of the absorption profile, and this parameter will affect the terminal wind speed. Generally, we aimed to fit the full blue edge, including this gradient. This means that the wind turbulence included in the SEI model adds to the overall wind speed and $v_{\infty}$ is reduced to compensate for this. As a result, $v_{\infty}$(SEI) is usually lower, causing the difference between $v_{\infty}$(SEI) and $v_{\infty}$(DM). Including this (often supersonic) turbulence in the SEI model essentially acts as broadening the profile function, but this has a similar effect on the emergent line profile as a more physically motivated simulation in terms of multiple non-monotonic velocity fields (shock structure), as suggested by \cite{Lucy1982} and \cite{Lucy1983} (see also \citealp{Puls1993}). The difference between $v_{\infty}$ as derived from $v_{\infty}$(DM) or from $v_{\infty}$(SEI) also depends on the strength of the line (which is proportional to the wind strength). If the strength is not too large (but large enough to saturate), the broadening by the large $v_{\rm{turb}}$ will not play a very strong role, and the difference will be small. However, when the strength is (very) large, the profile function can contribute also from regions that lie quite far away from line centre, and $v_{\infty}$(DM) becomes larger than $v_{\infty}$(SEI). This is somewhat different compared to a simulation using the wind-shock structure because in the latter case, the high-velocity regime is usually quite thin, so that its contribution remains more limited (unless the high-velocity interclump material is not as thin as usually adopted, see \citealp{Zsargo2008, Hawcroft2021, Brands2022}).
The SEI model is optimised using an optimisation routine (curve fit from the SciPy package, \citealp{SciPy}) based on the Levenberg-Marquardt evaluation criteria. The statistical error from this fitting routine is the $v_{\infty}$(SEI) uncertainty listed in Tables \ref{tbl:Results-LMC-sgiants} to \ref{tbl:Results-SMC-dwarfs}.

\section{Results} \label{sec: results}

\begin{figure}[H]
    \centering
    \includegraphics[scale=0.3]{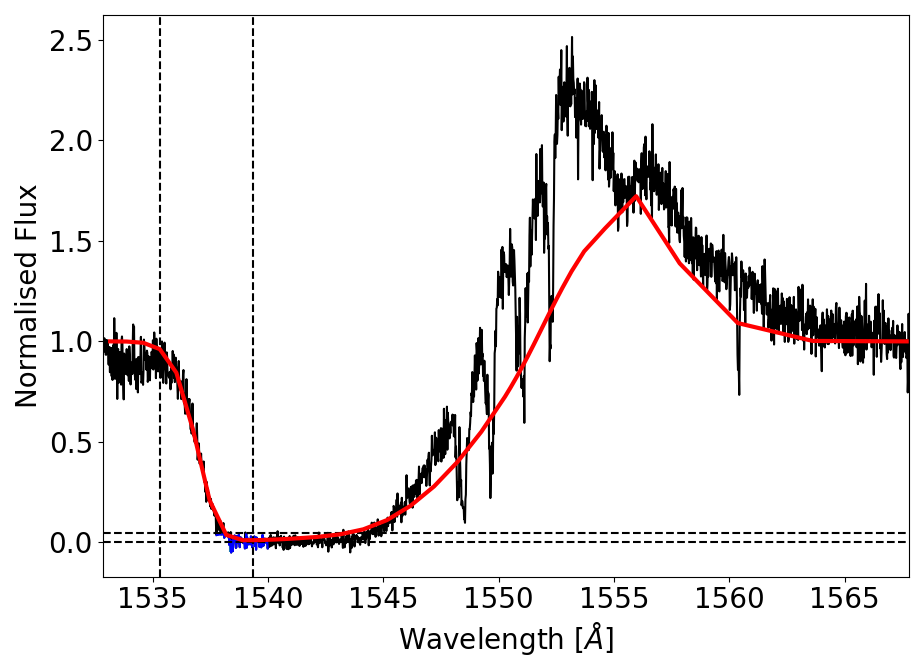}
    \includegraphics[scale=0.3]{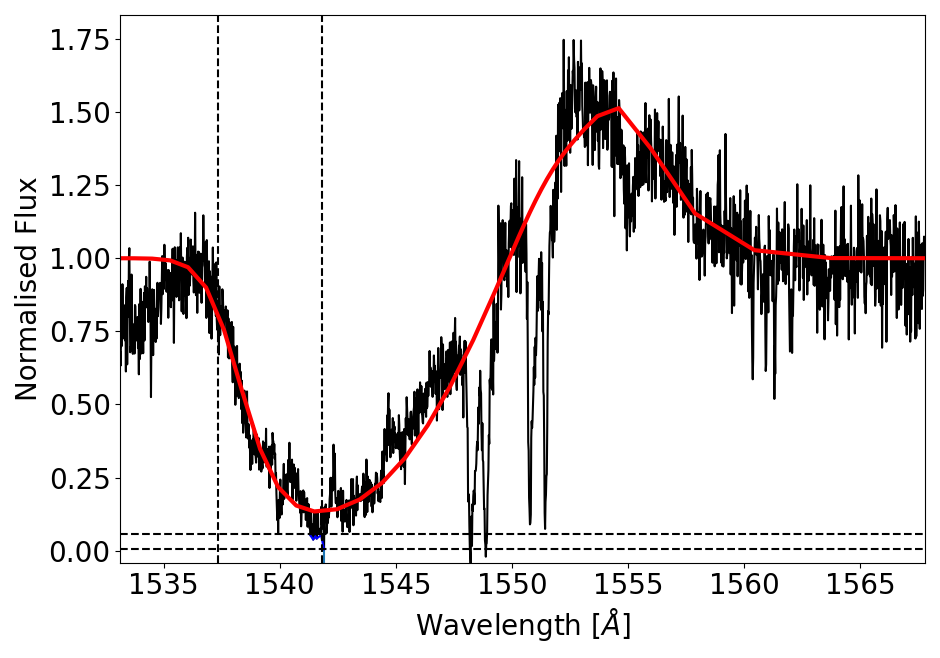}
    \includegraphics[scale=0.3]{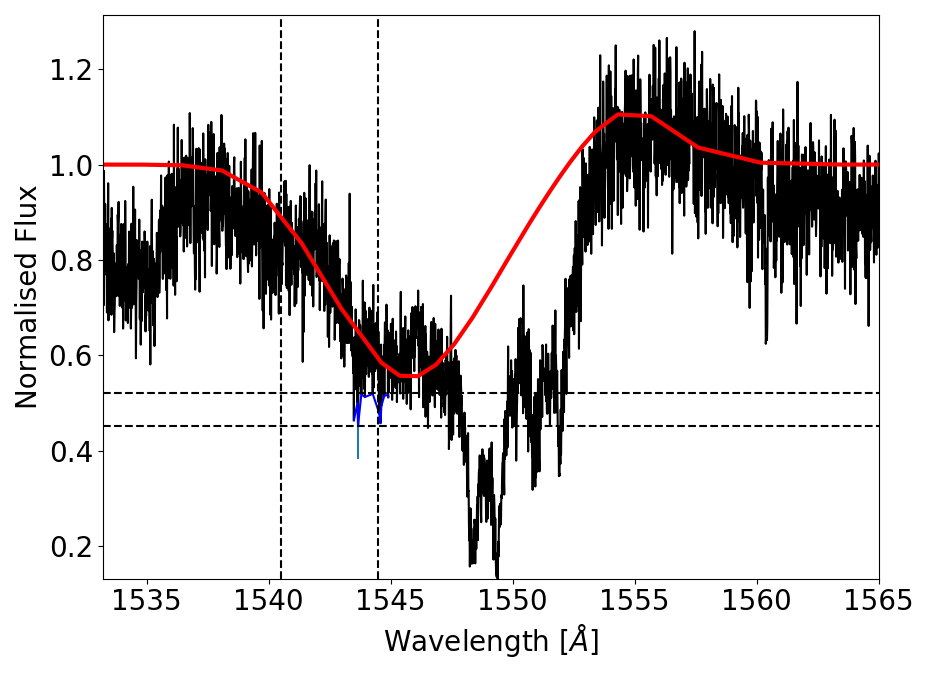}
        \includegraphics[scale=0.3]{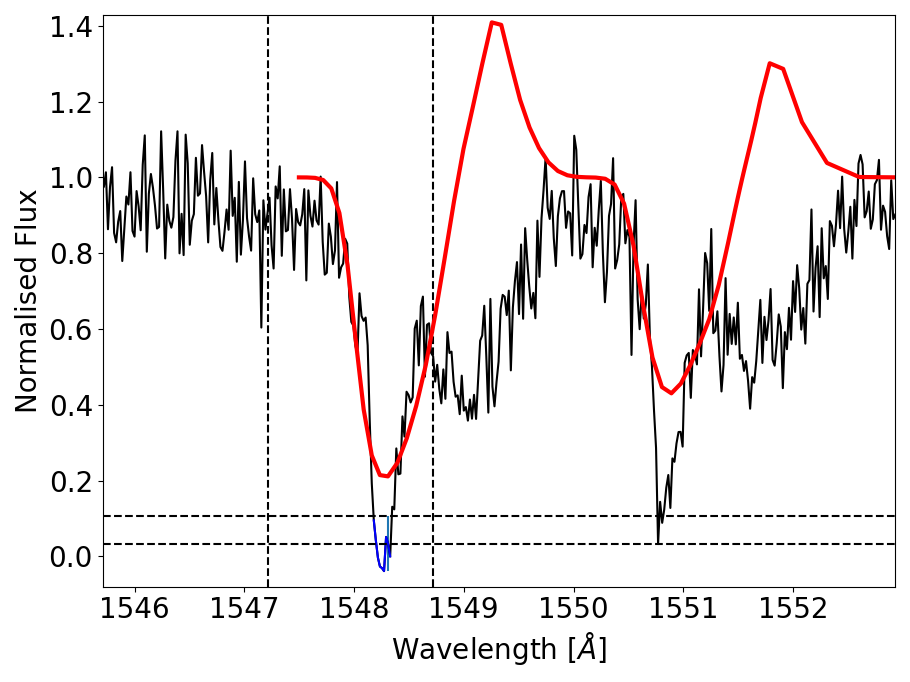}    \caption{Example of \ion{C}{iv} $\lambda\lambda$1548-1550 line profiles for different reliability categories from highest (top panel, rank i) to lowest (bottom panel, rank iv). Two profiles shown in this figure are from LMC stars (Sk --65$^{\circ}$ 47 and Sk --71$^{\circ}$ 19), and two are from SMC stars (AzV 26 and AzV 216) from the highest to the lowest panel: Sk --65$^{\circ}$ 47(O4If), AzV 26 (O6If), Sk --71$^{\circ}$ 19 (O6III), and AzV 216 (B1III). Black lines are the observed spectra. Red lines are the best-fit SEI model. Vertical dashed black lines highlight the fitting region for the SEI model. Horizontal dashed black lines show the S/N limit on the minimum flux of 0. Blue lines are the minimum flux within the S/N. } 
    \label{fig: Profile-Green-LMC}
\end{figure}

In this section, we present our measured terminal wind speeds along with stellar parameters from the ULLYSES metadata. Some subjective quality assessment was made during this analysis in order to eliminate stars for which the measurement of the terminal wind speed is unreliable because we did not observe the true terminal wind speed. The profiles are ranked by reliability to determine a terminal wind speed, with a poorer ranking given to stars that either have relatively low S/N spectra in the region of interest or that lack significant wind signatures. The categories are as follows: \begin{enumerate}[i]
  \item  Best quality: Clear and extended (>1\AA) saturation in the C\,{\sc iv}\,$\lambda \lambda$1548-1551 profile.
  \item Good quality: Strong wind signature, showing either a small saturation region or a significant absorption trough, >1\AA{} of minimum flux is below the level of half the continuum.
  \item Poor quality: Weak wind signature, trough is not significantly separated from continuum (minimum flux is above half of continuum average).
  \item Worst quality: No estimate of the wind speed can be made either due to data quality or lack of wind signature.
\end{enumerate} Examples of line profiles for each category are shown in Figure \ref{fig: Profile-Green-LMC}.

In the end, the sample yields 37 stars in the LMC and 16 in the SMC with saturated profiles fulfilling our highest-quality measurement criteria (rank i). This sub-sample was used to measure robust trends with escape speed, temperature, and metallicity, which are presented in Sections \ref{sec: vinf-vesc} to \ref{sec: vinf-z}. The larger sample is also considered and discussed in Appendix \ref{sec: low qual}. 

\begin{figure}[H]
    \centering
    \includegraphics[scale=0.2]{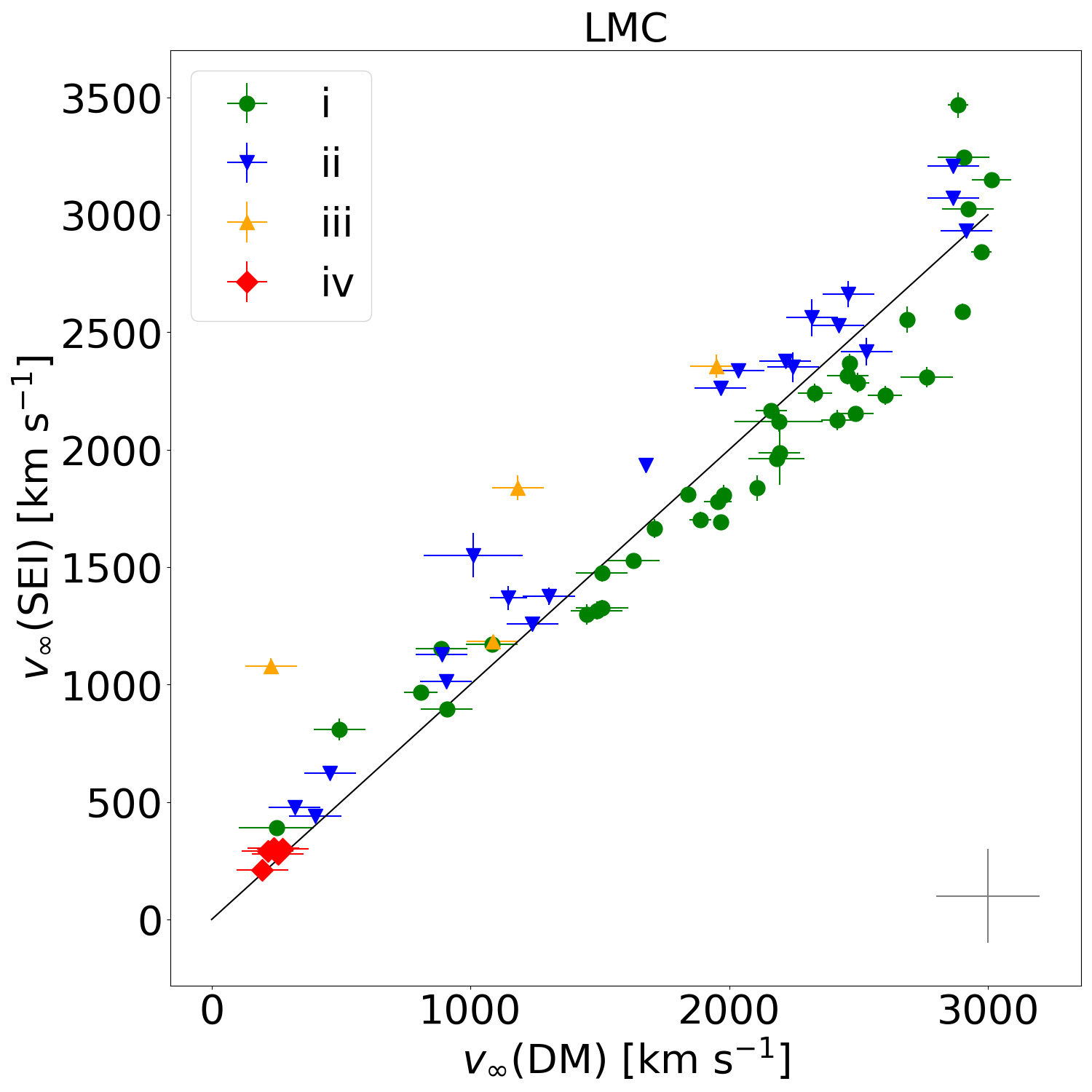}
    \includegraphics[scale=0.2]{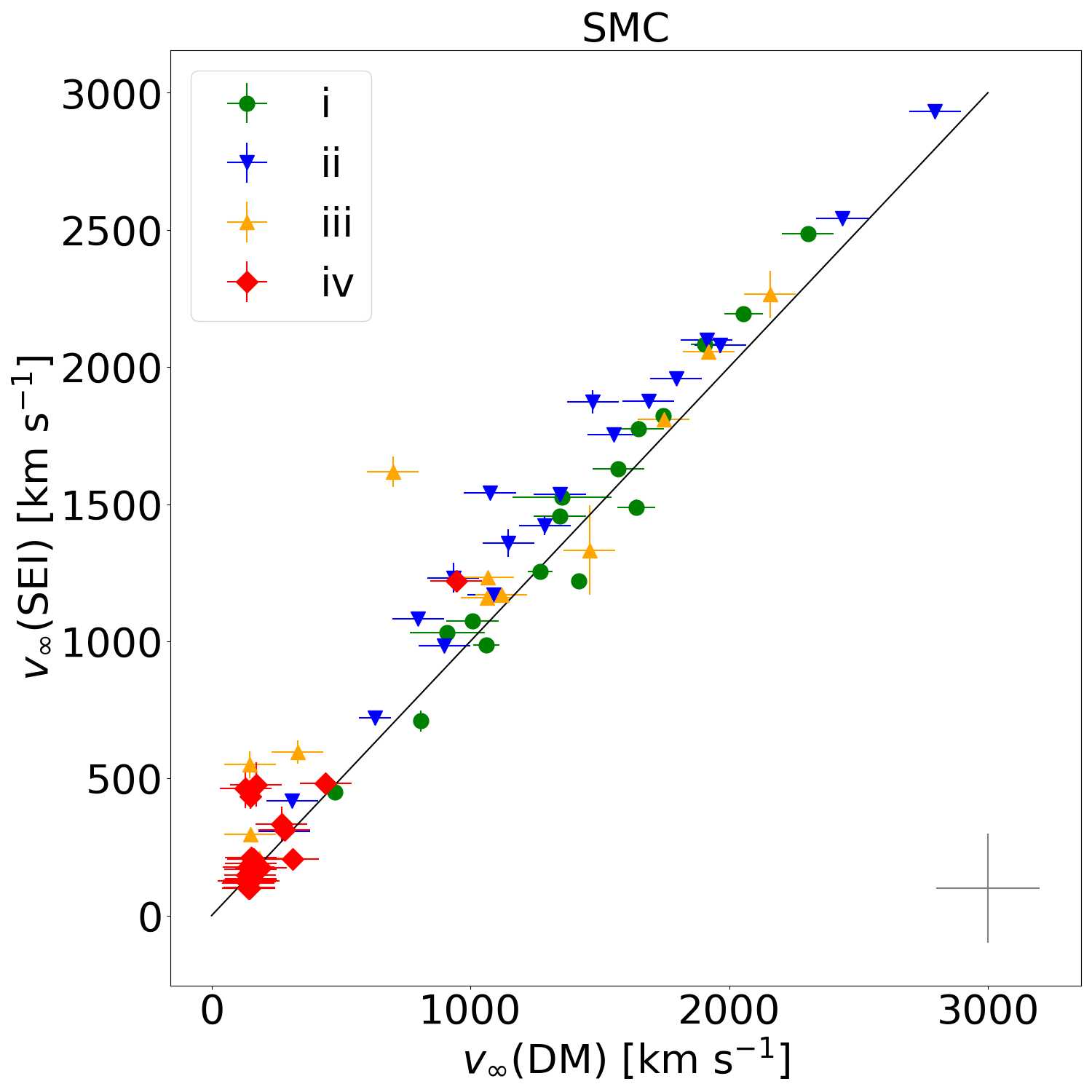}
    \caption{Comparison between terminal wind speed found using direct measurement $v_{\infty}$(DM), against that found using the SEI method $v_{\infty}$(SEI). \textit{Upper panel:} LMC sample. \textit{Lower panel:} SMC sample. Symbol shapes and colours indicate different quality categories from best (i) to worst (iv) (see the legend and Sect. \ref{sec: results}). The solid black line is a one-to-one relation between $v_{\infty}$(DM) and $v_{\infty}$(SEI). Conservative uncertainties of 200 $\rm{km}\, \rm{s^{-1}}$ on each measurement are visualised through the grey cross in the lower right corner of each panel.} 
    \label{fig: method_check}
\end{figure}

We list all measurements of terminal wind speed in Tables \ref{tbl:Results-LMC-sgiants} to \ref{tbl:Results-SMC-dwarfs}. When the terminal wind speeds are both measured directly (denoted by $v_{\infty}$(DM)) and using SEI fitting ($v_{\infty}$(SEI)), they generally agree well with each other (Fig. \ref{fig: method_check}). In terminal wind speed measurements of saturated profiles (best-quality rank, green circles in Fig. \ref{fig: method_check}), we find a systematic offset between $v_{\infty}$(SEI) and $v_{\infty}$(DM) because the addition of turbulence reduces $v_{\infty}$(SEI). However, this offset shift occurs more often in the LMC, which may be evidence of the effect discussed in Sect. \ref{sec: sei}, where in the LMC (higher wind densities), the majority of the saturated lines (in green) provide a $v_{\infty}$(DM) > $v_{\infty}$(SEI) because of a non-negligible influence of the high-velocity material. This effect would more or less vanish in the SMC because the high-velocity components affect the SMC less because of the lower wind densities. An opposing offset is present in measurements of the maximum observable wind speed in unsaturated profiles (lower quality rank, e.g. blue triangles in Fig. \ref{fig: method_check}); there are more stochastic discrepancies between the two methods on a case-by-case basis, with indeed some very large discrepancies in lower-quality measurements. For some stars, the point of minimum flux in the profile may be far enough from the blue edge that even within the S/N limit, the edge of the profile is not included in a direct measurement. This means that the offset occurs in the opposite direction ($v_{\infty}$(SEI) > $v_{\infty}$(DM)), as the SEI method attempts to fit the blue edge, which is excluded by direct measurement. The profile of AV26, shown in the second panel of Fig. \ref{fig: Profile-Green-LMC}, is an example of this, and the overall trend is clearest in the blue triangles in Fig. \ref{fig: method_check}.

Statistical errors on $v_{\infty}$(SEI) come from our optimisation routine. A more realistic precision for $v_{\infty}$(SEI) is to consider our values accurate to within 100 or 200 $\rm{km}\, \rm{s^{-1}}$ to account for the effect of wind turbulence on the measurement of $v_{\infty}$(SEI). As discussed in Sect. \ref{sec: sei}, the stronger the effect of turbulence, the larger the uncertainty on $v_{\infty}$.

\section{Discussion} \label{sec: discussion}

In this section, we explore the trends of the terminal wind speed with fundamental stellar parameters. The terminal wind speeds used here are those measured using the SEI method. We find that the differences between terminal wind speeds obtained using SEI or DM are not significant or systematic enough to affect these trends or conclusions. We also compare these trends with previous observational studies and applicable theoretical predictions. 

The linear trends were assessed with orthogonal distance regression (ODR), taking the errors on both variables into account, with an optimisation routine from the SciPy package applied to the relevant function \citep{SciPy}. For this discussion, we focus on the stars that fulfill our highest-quality measurement criteria. Further discussion is included in Appendix \ref{sec: low qual}. We expect some intrinsic scatter on the linear relations due to variability in the winds resulting from the line-deshadowing instability. There will also be scatter, especially at low Z, when different combinations of CNO abundances are accounted for because the outer wind is driven by only a few dozen metal lines (e.g. C, N, and O; see, e.g. \citealp{Krticka2006, Puls2008}).

\subsection{$v_{\infty} \:$ vs $\: v_{\rm{esc}}$} \label{sec: vinf-vesc}

\begin{figure}[t!]
    \centering
    \includegraphics[scale=0.25]{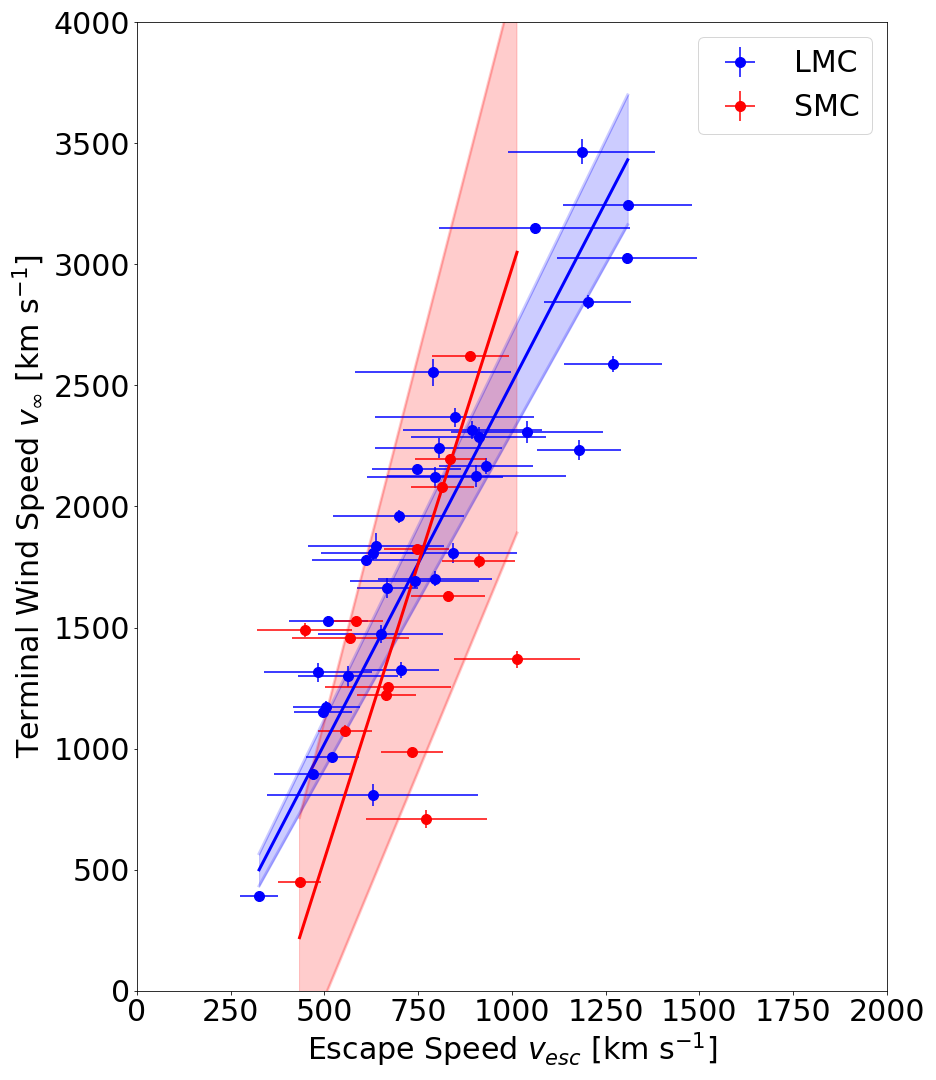}
    \caption{Terminal wind speed vs escape speed for LMC and SMC stars, shown in blue and red, respectively. Terminal wind speeds are measured by fitting SEI models to spectra, and the effective temperature estimates are acquired either through spectral type calibrations or from the literature. Escape speeds are estimated using evolutionary models computed with BONNSAI, given input stellar parameters from the literature or spectral type calibrations.} 
    \label{fig: Vinf-Vesc}
\end{figure}

\begin{figure}[t!]
    \centering
    \includegraphics[scale=0.25]{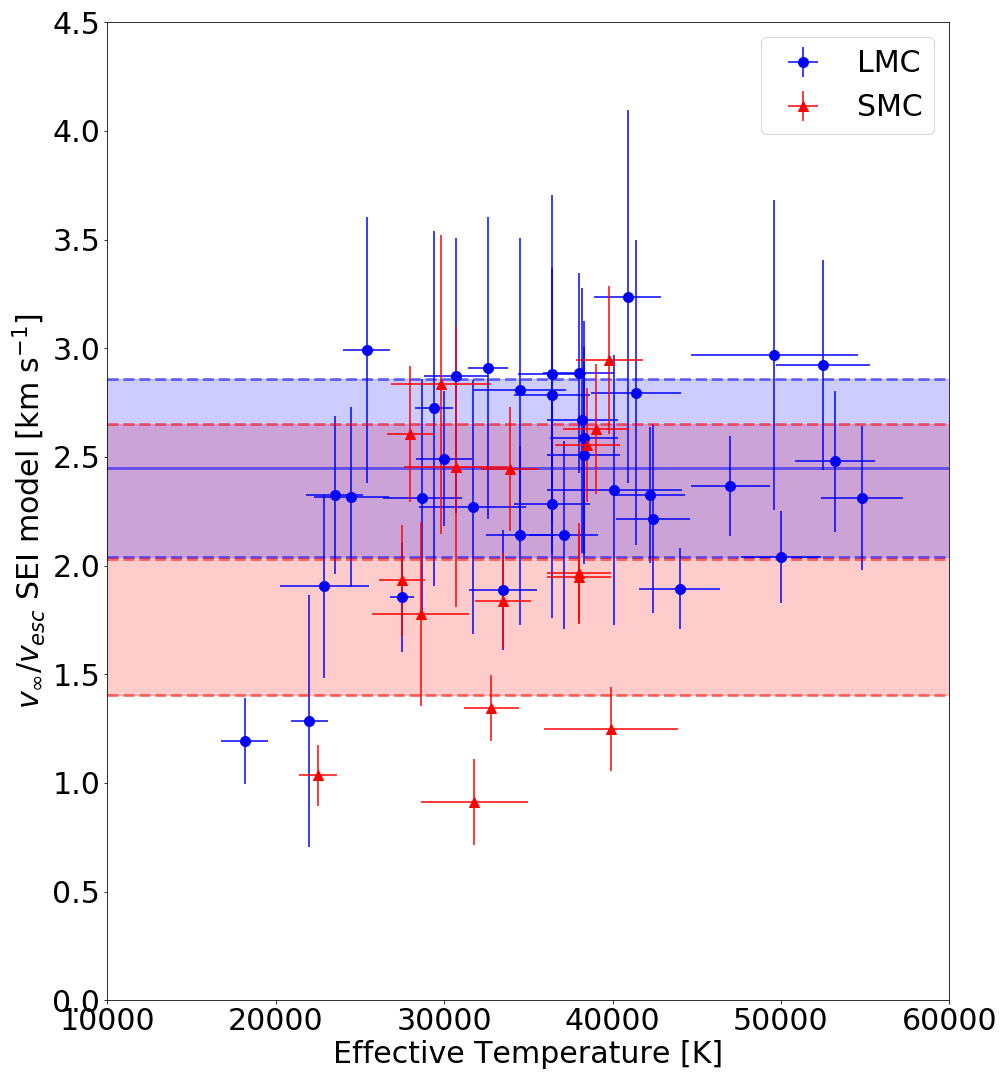}
    \caption{Ratio of the terminal wind speed to the escape speed vs effective temperature for LMC and SMC stars, shown in blue and red, respectively. The terminal wind speeds measured by fitting SEI models to spectra, and the effective temperature estimates are acquired either through spectral type calibrations or from literature.} 
    \label{fig: Teff-Vratio-LMC}
\end{figure}

The terminal wind speed has been predicted to follow strong relations with the surface escape velocity. A straightforward scaling with escape speed is predicted from radiation line driven wind theory, of the form \mbox{$v_{\infty} = 2.25(\alpha/(1-\alpha))v_{\rm{esc}}$} , where $\alpha$ is the main line force multiplier parameter \citep{Kudritzki1989, Kudritzki2000}. The most commonly implemented scaling is a constant $v_{\infty} = 2.65v_{\rm{esc}}$ that comes from the combination of a number of observational studies \citep{Howarth1989, Prinja1990, Lamers1995, Howarth1997, Prinja1998, Puls1996, Kudritzki1999}, compiled in \cite{Kudritzki2000}. This scaling was found to apply only at temperatures above 21 kK, and to decrease at lower temperatures. Temperature effects are discussed in the next sub-section. This decrease initially appeared as a step down to another constant \citep{Lamers1995}, and was interpreted as a bistability \citep{Pauldrach1990}, attributed to a change in ionisation balance \citep{Vink1999}. Further empirical studies suggest a more gradual decrease \citep{Evans2004, Crowther2006}. \cite{Markova2008} compiled a number of observational studies, along with new measurements, to investigate this trend, and again found a gradual decrease down to roughly 13 kK. Below this point, any estimates of the terminal wind speed are highly uncertain due to a lack of observations of necessary diagnostics. Our sample contains only a few objects with an effective temperature lower than 25 kK. Thus, we cannot contribute to this discussion here.

The relation we find between the terminal wind speed and the escape speed for the LMC is 

\begin{equation}
v_{\infty} = (3.0 \pm 0.2)v_{\rm{esc}} - (470 \pm 150), RMS=351 \: \rm{km\,s}^{-1}.
\end{equation}

\noindent This gradient is close to the relation of \cite{Kudritzki2000}, based on empirical studies of Galactic objects, and with relatively low scatter. For the SMC, we have 

\begin{equation}
v_{\infty} = (4.9 \pm 1.1)v_{\rm{esc}} - (1900 \pm 780), RMS=700 \: \rm{km\,s}^{-1},
\end{equation}

\noindent although here the scatter is large. Relations between terminal wind speed and surface escape speed in the LMC and SMC are shown in Fig. \ref{fig: Vinf-Vesc}. 

We find average values of $v_{\infty}/v_{esc}(LMC) = 2.4 \pm 0.4$ and $v_{\infty}/v_{esc}(SMC) = 2.0 \pm 0.6$. The ratio of wind speeds is shown as a function of effective temperature in Fig. \ref{fig: Teff-Vratio-LMC}. These values agree well with the empirical Galactic ratio of terminal wind to escape speed from \citealp{Kudritzki2000}, corrected for our predicted reduction in wind speed with metallicity (discussed further in Sect. \ref{sec: vinf-z}). \cite{Krticka2006} and \cite{Evans2004} find $v_{\infty}/v_{esc} = 2.3$ and $v_{\infty}/v_{esc} = 2.63$ for samples at SMC metallicity, respectively. The ratio from \cite{Evans2004} is determined empirically, while \cite{Krticka2006} used a combination of theoretical predictions and empirical determinations. \cite{Markova2008} find $v_{\infty}/v_{esc} = 3.3 \pm 0.7$ from stellar atmosphere fitting of a sample of Galactic supergiants with effective temperatures above 23 kK. Considering the large scatter and exclusion of lower values in the region between 21 kK and 23 kK, this is not a large disagreement with other estimates. \cite{Bjorklund2021} predicted a ratio higher than empirical studies with large scatter of $v_{\infty}/v_{esc} = 3.3 \pm 0.8$ for their Galactic models. These authors also find a strong trend of a decreasing ratio with increasing luminosity. \cite{Muijres2012} predicted a decreasing ratio with increasing temperature in their models with averages of $v_{\infty}/v_{esc}$ in a range from 3.1 to 3.6, depending on the modelling technique. However, theoretical predictions of $v_{\infty}/v_{esc}$ only extend to lower temperature limits of around 28 kK \citep{Muijres2012, Bjorklund2021}. If there is a flattening of the empirical $v_{\infty}/v_{esc}$ relation above this temperature (which could be argued from Fig. \ref{fig: Teff-Vratio-LMC}), then the theoretical and observational temperature dependences could converge at high temperatures and diverge mainly at lower temperatures. To resolve this, more measurements and predictions of $v_{\infty}$ for stars with temperatures below 28 kK are required, which could be tested against the predictions for B supergiants from \cite{Krticka2021}.

It is unclear whether the theoretical predictions of $v_{\infty}/v_{esc}$ (e.g. \citealp{Bjorklund2021}, \citealp{Muijres2012}) are truly larger than observational findings (e.g. \citealp{Kudritzki2000}) because the parameter ranges of the two techniques do not fully overlap. However, it is possible that $v_{\infty}$ is overestimated in models because physical processes that serve to alter the acceleration of the outer wind are underestimated. Some studies suggested that low-density winds are hotter on average because radiative cooling is inefficient, such that metals are present in the bulk wind at higher ionisation states \citep{Lucy2012, Lagae2021}. This would reduce the radiation force and lower the terminal wind speeds in stellar wind models. Moreover, the opposite effect would occur observationally, and we would not see strong wind lines in the typical ionisation states. Therefore, we would find (artificially) low terminal wind speeds in the usual diagnostics. This could explain a number of our quality rank iv measurements of $v_{\infty}$. It may also be that the evolutionary masses used in empirical studies are systematically overestimated \citep{Herrero1992}.

\subsection{$v_{\infty} \:$ versus $\: T_{\rm{eff}}$} \label{sec: vinf-teff}

\begin{figure}[t!]
    \centering
    \includegraphics[scale=0.25]{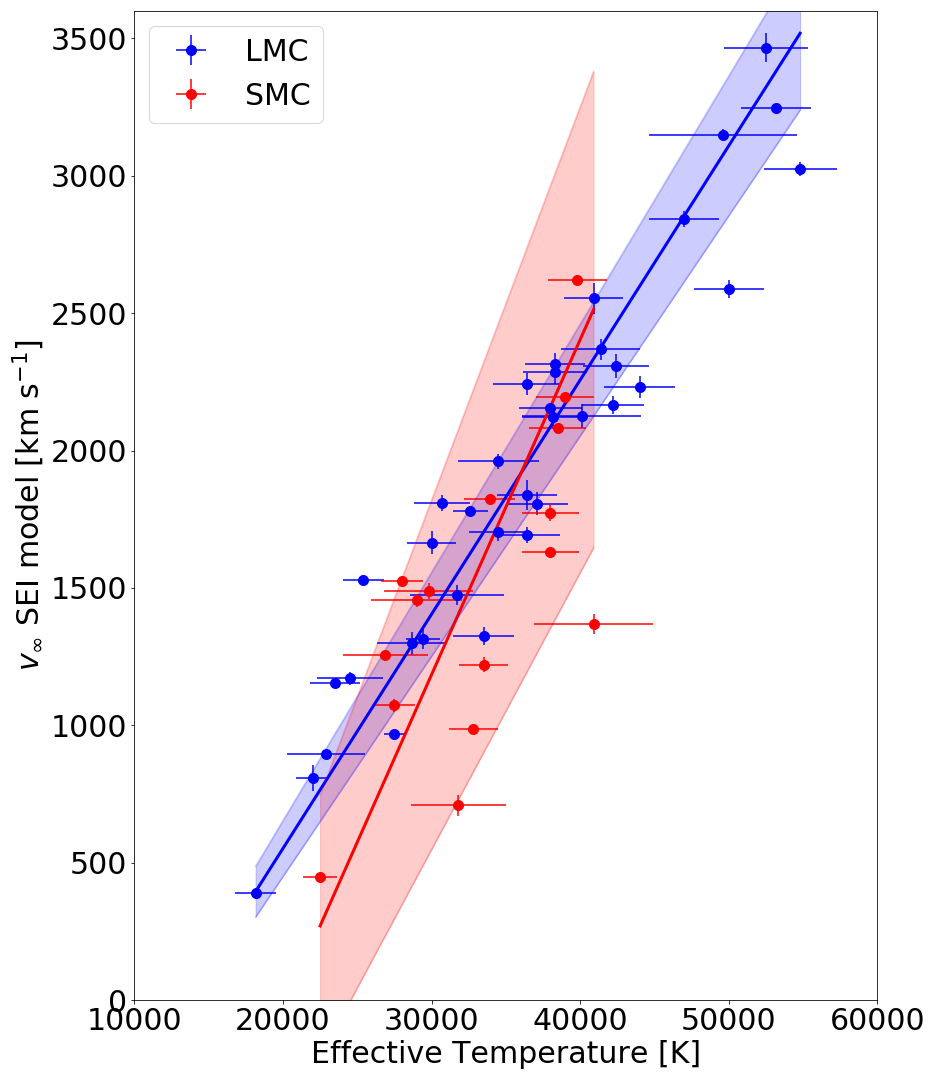}
    \caption{Terminal wind speed vs effective temperature for LMC and SMC stars, shown in blue and red, respectively. The terminal wind speeds are measured by fitting SEI models to spectra, and the effective temperature estimates are acquired either through spectral type calibrations or from the literature.} 
    \label{fig: Vinf-Teff-SEI}
\end{figure}

\begin{figure}[t!]
    \centering
    \includegraphics[scale=0.25]{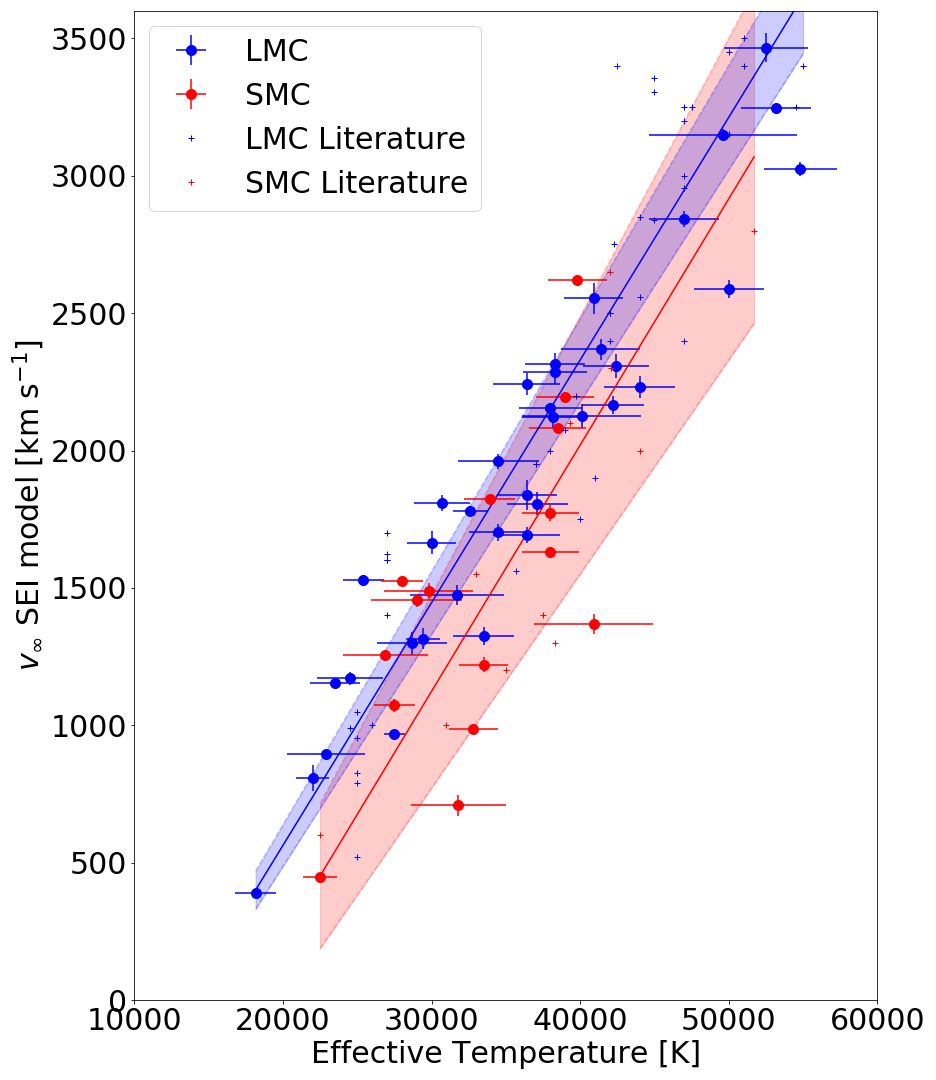}
    \caption{As Fig. \ref{fig: Vinf-Teff-SEI}, but including high-quality measurements of $v_{\infty}$ previously reported in the literature.} 
    \label{fig: Vinf-Teff-SEI-lit}
\end{figure}

\begin{table}
\centering
\caption{ODR fit coefficients to relations between $v_{\infty}$ and $T_{\rm{eff}}$ as described in Eq. \ref{eqn: teff}. }
\label{tbl:Results-Fits}
\vspace{2mm}
\begin{tabular}{lcccc}
\hline
\hline

Region & a & b & RMS & Sample  \cr
  & [$10^{-2}$] & [$\rm{km}\, \rm{s^{-1}}$] & [$\rm{km}\, \rm{s^{-1}}$] &  \\
\hline
LMC     &       $8.5 \pm 0.5$   &       $1150 \pm 170$  &       237      &  1  \\
SMC     &       $12.2 \pm 2.1$  &       $2470 \pm 680$  &       477      &  1  \\
LMC     &       $8.8 \pm 0.4$   &       $1200 \pm 150$  &       312      &  1,2  \\
SMC     &       $8.9 \pm 1.1$   &       $1560 \pm 420$  &       372      &  1,2  \\
GAL     &       $10.2 \pm 0.3$  &       $1300 \pm 100$  &       277      &  2  \\

\hline
\\              
\end{tabular}
\begin{tablenotes}
\item{The column 'sample' indicates whether the fit is based on data from this paper (1) or from the literature (2), or both (1,2). The column 'RMS' refers to the root mean square error.}
\end{tablenotes}
\end{table}

We find a clear trend of increasing terminal wind speed with effective temperature. This is shown in Fig. \ref{fig: Vinf-Teff-SEI} with linear fits of the form

\begin{equation} \label{eqn: teff}
v_{\infty} = aT_{\rm{eff}} - b.
\end{equation}

\noindent The fit coefficients for different sub-samples are presented in Table \ref{tbl:Results-Fits}. This general trend is well established, but it is most commonly tied to the escape speed, with which there is also a clear trend, but with slightly larger scatter. Considering the RMS of fits against both parameters, and that the known uncertainties in determining stellar masses and radii are larger than those on temperature, perhaps a more accurate empirical prediction of the terminal wind speed can be made using effective temperature rather than escape speed.

For the SMC, the trend with effective temperature shows a larger dispersion, spanning a limited range of 22-40 kK in temperature. However, the sample comprises only 16 stars, less than half of the equivalent LMC sample. As a result, the uncertainty is much larger than that found for the LMC. We can double the sample size by including stars from ULLYSES with slightly lower quality measurements (see Appendix \ref{sec: low qual}). However, only a few points are added above 40 kK when these stars are included, and we do not measure the true terminal wind speed, only a lower limit. We also find different slopes with much higher RMS when we include lower quality measurements in the LMC sample, which suggests that the trends found using these measurements are less reliable. 

The temperature coverage in the SMC can be extended by adding equally high-quality measurements from literature samples compiled in \cite{Garcia2014}, and additional measurements from \cite{Bouret2021}. The high temperature coverage is provided by essentially one star (NGC346 MPG 355) that is included in the ULLYSES sample, but does not show saturation in the C\,{\sc iv}\,$\lambda \lambda$1548-1551 and so is not included in our highest-quality rank sub-sample. However, this star does show saturation in the N\,{\sc v}\,$\lambda \lambda$1239-1243 profile, which is fitted in \cite{Bouret2021}, therefore we chose to include this literature measurement. Furthermore, the measured terminal wind speeds for this star in this work and \cite{Bouret2021} agree to within 150 km\,s$^{-1}$. Adding literature values from \cite{Garcia2014} to the LMC sample does not significantly impact the slope. This suggests that adding literature values for the SMC should serve only to reduce the uncertainty on our trend as the literature values add coverage at high effective temperature. When considering the literature sample of the SMC, we find a gradient similar to the LMC. The effect of including literature values is shown in Fig. \ref{fig: Vinf-Teff-SEI-lit}. We note that the majority of the stars included in the highest-quality ranks, and so used to determine these relations, are supergiants. For example, in Fig. \ref{fig: Vinf-Teff-SEI}, the LMC sample comprises 70\% luminosity class I stars, 20\% class II-III and 10\% class V. This means that these results mainly apply to stars with denser winds. 

We may find a physical motivation of the dependence of the terminal wind speed on temperature by considering the mass-luminosity relation. As discussed in Sect. \ref{sec: vinf-vesc}, we expect $v_{\infty} \propto v_{\rm{esc}}$ and $v_{\rm{esc}} \propto ( M / R )^{0.5}$ , as shown in Eqn. \ref{vesc-eqn}. Using these relations and a mass-luminosity relation of the nature $L \propto M^{\alpha}$, we find $v_{\infty} \propto v_{\rm{esc}} \propto ( L^{1/\alpha} / R )^{0.5}$. If we then substitute the luminosity for the Stefan-Boltzmann law ($L \propto R^{2}T_{\rm{eff}}^{4}$), we find that $v_{\infty} \propto ( (R^{2}T^{4})^{1/\alpha} / R )^{0.5}$. This results in a relation between terminal wind speed and temperature that becomes linear ($v_{\infty} \propto T_{\rm{eff}} $) when $\alpha \sim 2$. This offers some qualitative evidence for a relation of the terminal wind speed and temperature, although it is thought that the $\alpha$ exponent is slightly larger for more massive stars (\citealp{Eker2015} find $\alpha \sim 2.7$ for main-sequence stars between 7 and 32 $M_{\odot}$). 

\subsection{$v_{\infty} \:$ versus $\: Z$} \label{sec: vinf-z}

\cite{Leitherer1992} offered a theoretical power-law index for the terminal wind speed of the form $v_{\infty} \propto Z^{0.13}$ considering radiative transfer models of massive stars ($>15\,M_{\odot{}}$) covering a wide metallicity range $0.01\,Z_{\odot} \leq Z < 3.0\,Z_{\odot}$. This dependence becomes $v_{\infty} \propto Z^{0.15}$ when the $Z$ range is restricted to $0.1\,Z_{\odot} \leq Z < 3.0\,Z_{\odot}$. Further theoretical predictions for $v_{\infty}$ come from \cite{Krticka2006}, \cite{VinkSander2021} and \cite{Bjorklund2021}. \cite{Krticka2006} offered the relation $v_{\infty} \sim Z^{0.06}$, although this prediction comes from comparing around 20 theoretical models at two metallicity points, $Z=0.2\,Z_{\odot}$ and $Z=0.3\,Z_{\odot}$. Using the Monte Carlo approach from \cite{Muller2008} to predict terminal velocities and mass-loss rates, \cite{VinkSander2021} obtained $v_{\infty} \propto Z^{0.19}$ for $0.033\,Z_{\odot} \leq Z < 3\,Z_{\odot}$. \cite{Bjorklund2021} predicted $v_{\infty} \propto Z^{-0.10 \pm 0.18}$. However, these authors also discussed issues with their predictions of terminal wind speed. We note further that these are average values. \cite{Bjorklund2021} reported that the $v_{\infty}/v_{esc}$ dependence is affected by the luminosity of stars considered, which also affects the metallicity dependence. The metallicity dependence of $v_{\infty}$ has been investigated empirically by \cite{Garcia2014}. Their study included stars in the Galaxy, M31, M33, LMC, and SMC by compiling a number of studies that were then compared and extended further to IC 1613, which is suggested to have a very low metallicity ($\leq 0.1\,Z_{\odot}$). Altogether, these authors suggest a more complex metallicity dependence. For this study, we did not include M31, M33, or IC 1613 because the sample sizes are small, but we searched through the literature for $v_{\infty}$ measurements at Galactic metallicity.

We found a suitable Galactic sample from the stars with saturated wind line profiles in \cite{Prinja1990}. In that paper, spectral types were determined or taken from previous references, and stellar parameters come from spectral type calibrations in \cite{Howarth1989}. We used the measurements of $v_{\infty}$ from \cite{Prinja1990}, but we used the stellar parameters from \cite{Holgado2020}. Approximately 95\% of the stars with saturated C\,{\sc iv}\,$\lambda \lambda$1548-1551 profiles from \cite{Prinja1990} are included in \cite{Holgado2020}. The stellar parameters of 70\%\ of these stars were determined with quantitative spectroscopy, the others are identified as double-lined spectroscopic binaries, and \cite{Holgado2020} therefore did not present spectroscopic fits to these stars. We use this sub-sample of 66 stars with measurements of $v_{\infty}$ from \cite{Prinja1990} and effective temperatures from \cite{Holgado2020} in this section. In Appendix \ref{App-Galaxy} we use the full \cite{Prinja1990} sample by updating spectral types using the Galactic O Star Spectroscopic Survey (GOSSS; \citealp{Sota2011, Sota2014, Apellaniz2016} and using these types to determine stellar parameters from the calibrations of \cite{Martins2005}. We do not find a significant difference in the terminal wind speed with metallicity dependence for either Galactic sub-sample. There are also a number of B supergiants in the Prinja sample; these stars have no equivalent catalogue, therefore we confirmed the spectral types against various literature sources and updated them when the spectral type was revised in the subsequent literature. To determine stellar parameters for these stars, we used the spectral type calibration from \cite{McEvoy2015}, who used \cite{Crowther2006} in the Galaxy. For the few B giants and dwarfs, we used the calibration of \cite{Dufton2006}. The caveat then is that the effective temperatures for B stars in the relation still come from spectral type calibrations and could cause a systematic offset between the temperatures of the B and O stars. The consistency between B supergiant temperature calibrations from \cite{Markova2008} (which agree very well with those from \citealp{Crowther2006}) and O star parameters, either from \cite{Martins2005} calibrations or \cite{Holgado2020}, has been quantified by \cite{Berlanas2018}. These authors concluded that the \cite{Holgado2020} parameters are preferred because the \cite{Martins2005} calibrations give low effective temperatures for late-O stars. This is also discussed in \cite{SimonDiaz2014}. Therefore, we also chose to use the O star parameters from \cite{Holgado2020} over the calibrations from \cite{Martins2005}. We did not separate the sample into luminosity class in this section, but the effect of doing this is minimal. Therefore, the breakdown by luminosity class is presented in Appendix \ref{lum-class}. The effect of including other Galactic samples \citep{Groenewegen1989, Crowther2006} is minimal and is presented in Appendix \ref{App-Galaxy}.

\begin{figure}[t!]
    \centering
    \includegraphics[scale=0.25]{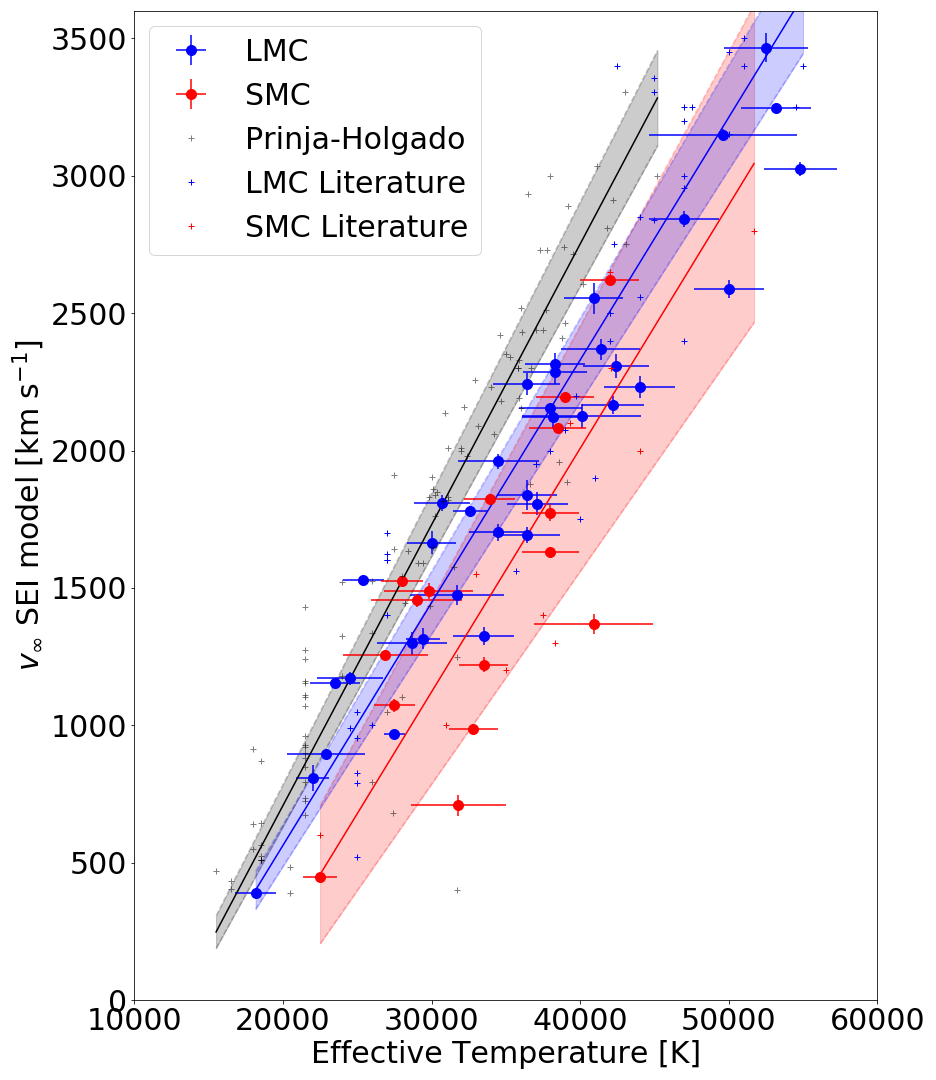}
    \caption{As Figure \ref{fig: Vinf-Teff-SEI}, but including the literature sample for Galactic OB stars to illustrate the metallicity dependence of the terminal wind speed.} 
    \label{fig: Vinf-Teff-Zdep}
\end{figure}

To quantify the relation between $v_{\infty}$ and metallicity, we combined the highest-quality measurements in the Galaxy, LMC, and SMC, along with the appropriate LMC and SMC literature measurements discussed in Sect. \ref{sec: vinf-teff}, and applied a multi-dimensional linear fit to $v_{\infty}$ as a function of $T_{\rm{eff}}$ and $Z$. This yielded 

\begin{equation} \label{eq:vinf-Z}
v_{\infty} = \left[ 9.2(\pm0.3)10^{-2}T_{\rm{eff}} - 1040(\pm100) \right] Z^{(0.22 \pm 0.03)},
\end{equation}

\noindent giving a final dependence of terminal wind speed on metallicity of the form $v_{\infty} \propto Z^{0.22\pm0.03}$. With an $RMS = 305$ km\,s$^{-1}$, we are able to reliably predict $v_{\infty}$ within a reasonable margin, offering an improvement on the uncertainty in stellar atmosphere and feedback modelling when $v_{\infty}$ diagnostics are unavailable. The relation between terminal wind speed and effective temperature for the Galaxy, LMC, and SMC is shown in Fig. \ref{fig: Vinf-Teff-Zdep}.

\section{Conclusions} \label{sec: conclusions}

We have empirically determined the terminal wind speeds for all OB type stars in the ULLYSES Data Release 3 sample. For stars with saturated \ion{C}{IV} $\lambda\lambda$1148-1150 line profiles, we directly measured the terminal wind speed. For those with unsaturated \ion{C}{IV} $\lambda\lambda$1148-1150 profiles, we measured the maximum observable wind speed, marking a lower limit on the terminal wind speed. We also estimated the terminal wind speed by performing a fitting routine using synthetic line profiles, computed with the SEI method. We compared the wind speeds to fundamental stellar properties, including effective temperatures collected and provided by the ULLYSES project. We also estimated a number of stellar properties that are not provided in the ULLYSES metadata using a comparison with single-star evolutionary tracks. Finally, we compared our findings with compiled literature studies using similar techniques to those implemented here. We note that these conclusions are primarily drawn from stars with relatively dense winds. An application of these relations to stars with lower density winds (e.g. very low mass-loss rates) is therefore debatable. Our results are listed below.

We find a linear relation between terminal wind speed and escape speed for current surface escape speeds. This trend is theoretically motivated by (semi-)analytical solutions for radiation-driven winds for OB stars \citep{Kudritzki1989}.

The average values of the ratio of the terminal wind speed to the surface escape speed are $v_{\infty}/v_{esc} = 2.4\pm0.4$ in the LMC and $v_{\infty}/v_{esc} = 2.0\pm0.6$ in the SMC. These values agree with previous empirical estimates above 21 kK \citep{Kudritzki2000} when they are corrected for a metallicity dependence $\sim Z^{0.2}$. However, the scatter around these averages is large, with $RMS = 278$ km\,s$^{-1}$ in the LMC and $RMS = 498$ km\,s$^{-1}$ in the SMC. We are unable to comment on the dependence of this ratio for temperatures below 21 kK because there are only a few stars in this sample with a low effective temperature.

We also find a trend of the terminal wind speed with the temperature, showing less scatter than the trend of the terminal wind speed with the escape speed. In the LMC, we find a dispersion on the trend with the temperature of $RMS = 237$ km\,s$^{-1}$, compared to $RMS = 351$ km\,s$^{-1}$ with the escape speed. In the SMC, we find $RMS = 477$ km\,s$^{-1}$ on the temperature trend, and $RMS = 700$ km\,s$^{-1}$ with the escape speed. Altogether, this suggests that we obtain the most reliable estimation of the terminal wind speed from the effective temperature.

Finally, we find a trend of the terminal wind speed with metallicity of $v_{\infty} \propto Z^{0.22 \pm 0.03}$, with $RMS = 305$ km\,s$^{-1}$. This is somewhat steeper than the theoretical prediction of $v_{\infty} \propto Z^{0.13}$ from \cite{Leitherer1992}, but agrees with the recent calculations by \cite{VinkSander2021}. 

\begin{acknowledgements}
We would like to thank the referee, Sergio Sim\'{o}n-D\'{i}az, for their helpful comments which improved the content and clarity of this paper. We thank Fabian Schneider for assistance with BONNSAI. This project has received funding from the  KU Leuven Research Council (grant C16/17/007: MAESTRO), the FWO through a FWO junior postdoctoral fellowship (No. 12ZY520N) as well as the European Space Agency (ESA) through the Belgian Federal Science Policy Office (BELSPO). L.M. thanks the European Space Agency (ESA) and the Belgian Federal Science Policy Office (BELSPO) for their support in the framework of the PRODEX Programme. JMB and PAC are supported by the Science and Technology Facilities Council research grant ST/V000853/1 (PI. V. Dhillon). NDK is supported by the National Solar Observatory (NSO). NSO is managed by the Association of Universities for Research in Astronomy, Inc., and is funded by the National Science Foundation. B.K. gratefully acknowledges support from the Grant Agency of the Czech Republic (GA\v CR 22-34467S). The Astronomical Institute in Ond\v rejov is supported by the project RVO: 67985815. AACS is funded by the Deutsche Forschungsgemeinschaft (DFG, German Research Foundation) in the form of an Emmy Noether Research Group -- Project-ID 445674056 (SA4064/1-1, PI Sander). AACS acknowledges further support by the Federal Ministry of Education and Research (BMBF) and the Baden-Württemberg Ministry of Science as part of the Excellence Strategy of the German Federal and State Governments. ADU acknowledges support by NASA under award number 80GSFC21M0002. M.G and F.N. acknowledge funding by grants PID2019-105552RB-C41 and MDM-2017-0737-19-3 Unidad de Excelencia "María de Maeztu". A.H. acknowledges support from the Spanish Ministry of Science and Innovation (MICINN) through the Spanish State Research Agency through grants PID2021-122397NB-C221 and the Severo Ochoa Programe 2020-2023 (CEX2019-000920-S). NSL wishes to thank the National Sciences and Engineering Council of Canada (NSERC) for financial support. Based on observations obtained with the NASA/ESA Hubble Space Telescope, retrieved from the Mikulski Archive for Space Telescopes (MAST) at the Space Telescope Science Institute (STScI). STScI is operated by the Association of Universities for Research in Astronomy, Inc. under NASA contract NAS 5-26555. This research has made use of the SIMBAD database, operated at CDS, Strasbourg, France. Figures of fits to the \ion{C}{IV} $\lambda\lambda$1548-1550 profile of each star will be shared upon request to the corresponding author.
\end{acknowledgements}

%"M.G. acknowledges funding from grants PID2019-105552RB-C41 and MDM-2017-0737-19-3"

% Dylan "The National Solar Observatory (NSO) is operated by the Association of Universities for Research in Astronomy, Inc. (AURA), under cooperative agreement with the National Science Foundation."

% AACS acknowledges support by the Deutsche Forschungsgemeinschaft (DFG - German Research Foundation) in the form of an Emmy Noether Research Group (grant number SA4064/1-1, PI Sander) For my affiliation in the author list, please write (keeping the German names): Zentrum für Astronomie der Universit\"{a}t Heidelberg, Astronomisches Rechen-Institut, M\"{o}nchhofstr. 12-14, 69120 Heidelberg

\bibliographystyle{aa} % style aa.bst
\bibliography{terminal.bib}

\begin{appendix}

\section{Measured terminal wind speeds} \label{sec: tables}

Here we present the measured terminal wind speeds along with spectral types and stellar parameters as described in Sect. \ref{sec: results}.

\begin{landscape}

\thispagestyle{plain}

\begin{table}
\tiny
\caption{Stellar parameters and spectral types from ULLYSES metadata along with terminal wind speeds for supergiants in the LMC.  }
\label{tbl:Results-LMC-sgiants}

\begin{tabular}{lllllllllllllllllll}
\hline
\hline

ID & Sp. Type   & log$(\frac{L}{L_{\odot}})$ & $\Delta$L & $T_{\rm{eff}}$ & $\Delta T_{\rm{eff}}$ & $v_{\infty}$(DM) & $\Delta v$ & $v_{\infty}$(SEI) & $\Delta v$ & $v_{\rm{turb}}$ & $\Delta v$ & $M_{\rm{evol}}$ & $\Delta M_{\rm{evol}}$ & $v_\mathrm{esc}$ & $\Delta v_\mathrm{esc}$ & Q & Alias & Ref  \cr
  &   & [dex] &  & [kK] & & [$\rm{km}\rm{s^{-1}}$] & & [$\rm{km}\rm{s^{-1}}$] & & [$\rm{km}\rm{s^{-1}}$] & & [$M_{\odot}$] & & [$\rm{km}\rm{s^{-1}}$] & \\
\hline
Sk --67$^{\circ}$ 22    &       O2\,If*/WN5     &       5.8     &       0.1     &       50      &       2.4     &       2900    &       20      &       2590    &       30      &       430     &       30      &       61      &       8       &       1480    &       130     &       i       &       BAT99-12        &       18      \\
BAT99-105       &       O2\,If* &       6.56    &       0.1     &       47.3    &       2.4     &       2920    &       100     &       2930    &       30      &       300     &       20      &       153     &       43      &       1500    &       240     &       ii      &       Mk 42      &       19      \\
VFTS 482        &       O2.5\,If*/WN6   &       6.4     &       0.1     &       42.2    &       2.1     &       2420    &       100     &       2530    &       30      &       560     &       30      &       111     &       22      &       1220    &       160     &       ii      &       Mk 39, BAT99-99    &       19      \\
VFTS 180        &       O3\,If* &       5.92    &       0.1     &       42.2    &       2.1     &       2160    &       60      &       2170    &       30      &       180     &       20      &       57      &       8       &       1180    &       120     &       i       &       BAT99-93,\,ST92\,1-78   &       19      \\
Sk --65$^{\circ}$ 47    &       O4\,If  &       5.98    &       0.2     &       40.5    &       2.7     &       2180    &       110     &       1960    &       30      &       430     &       20      &       37      &       9       &       910     &       170     &       i       &               &       15      \\
Sk --71$^{\circ}$ 46    &       O4\,If  &       6.1     &       0.1     &       38      &       2.1     &       2490    &       70      &       2160    &       10      &       540     &       0       &       68      &       11      &       1030    &       120     &       i       &       NGC206 FS 214  &       23      \\
Sk --67$^{\circ}$ 167   &       O4\,Inf$^{+}$   &       5.87    &       0.2     &       40.1    &       4.0     &       2420    &       60      &       2130    &       40      &       530     &       50      &       47      &       11      &       1180    &       240     &       i       &               &       1       \\
ST92 4-18       &       O5\,If  &       5.61    &       0.2     &       38.3    &       2.0     &       2460    &       80      &       2320    &       40      &       280     &       30      &       34      &       6       &       1080    &       180     &       i       &       W61 4-4     &       1       \\
LMCE078-1       &       O6\,Ifc &       5.6     &       0.2     &       36.4    &       2.2     &       2330    &       70      &       2240    &       40      &       560     &       40      &       32      &       6       &       980     &       170     &       i       &               &       1       \\
Sk --67$^{\circ}$ 111   &       O6\,Ia(n)fpv    &       5.74    &       0.2     &       36.4    &       2.3     &       1970    &       20      &       1690    &       30      &       350     &       30      &       37      &       8       &       960     &       170     &       i       &               &       1       \\
Sk --70$^{\circ}$ 115   &       O6\,If  &       6.02    &       0.2     &       36.4    &       2.0     &       2110    &       10      &       1840    &       50      &       460     &       50      &       50      &       13      &       950     &       180     &       i       &       HDE 270145  &       1       \\
Sk --69$^{\circ}$ 104   &       O6\,Ib(f)       &       5.94    &       0.2     &       36.4    &       2.1     &       2460    &       100     &       2660    &       60      &       520     &       40      &       47      &       11      &       990     &       180     &       ii      &       HDE 269357  &       1       \\
Sk --65$^{\circ}$ 22    &       O6\,Iaf$^{+}$   &       5.86    &       0.1     &       33.5    &       2.0     &       1510    &       100     &       1330    &       30      &       330     &       30      &       48      &       6       &       900     &       100     &       i       &               &       2       \\
Sk --69$^{\circ}$ 50    &       O7(n)(f)p       &       5.5     &       0.2     &       36.1    &       2.0     &       1890    &       40      &       1700    &       30      &       420     &       30      &       28      &       5       &       930     &       150     &       i       &               &       1       \\
Sk --67$^{\circ}$ 168   &       O8\,I(f)p       &       5.84    &       0.2     &       32.6    &       1.2     &       1960    &       50      &       1780    &       20      &       380     &       20      &       39      &       10      &       830     &       150     &       i       &       HDE 269702  &       1       \\
LH 9-34 &       O8.5\,Iaf       &       5.67    &       0.2     &       31.7    &       3.2     &       1510    &       100     &       1470    &       40      &       370     &       30      &       32      &       7       &       830     &       170     &       i       &       PGMW\,1363, BI\,37  &       1       \\
Sk --67$^{\circ}$ 107   &       O9\,Ib(f)       &       5.61    &       0.2     &       30.7    &       1.9     &       1840    &       10      &       1810    &       30      &       420     &       30      &       30      &       7       &       780     &       140     &       i       &               &       1       \\
Sk --69$^{\circ}$ 279   &       O9.2\,Iaf       &       5.54    &       0.1     &       29.5    &       1.7     &       460     &       100     &       620     &       10      &       150     &       10      &       32      &       4       &       770     &       80      &       ii      &               &       25      \\
Sk --71$^{\circ}$ 41    &       O9.7\,Iab       &       5.5     &       0.1     &       30      &       1.7     &       1710    &       30      &       1660    &       40      &       360     &       40      &       30      &       3       &       780     &       80      &       i       &       NGC206 FS 134  &       23      \\
Sk --67$^{\circ}$ 5     &       O9.7\,Ib        &       6.04    &       0.2     &       29.4    &       1.2     &       1490    &       100     &       1320    &       40      &       330     &       40      &       49      &       14      &       750     &       140     &       i       &       HDE 268605  &       1       \\
VFTS 87 &       O9.7\,Ib-II     &       5.29    &       0.2     &       30.5    &       2.9     &       1010    &       190     &       1550    &       100     &       390     &       120     &       21      &       3       &       850     &       180     &       ii      &               &       22      \\
Sk --68$^{\circ}$ 135   &       ON9.7\,Ia$^{+}$ &       5.97    &       0.1     &       27.5    &       0.7     &       810     &       60      &       970     &       20      &       240     &       20      &       52      &       7       &       710     &       70      &       i       &       HDE 269896  &       3       \\
Sk --68$^{\circ}$ 52    &       B0\,Ia  &       5.76    &       0.1     &       24.5    &       2.2     &       1080    &       100     &       1170    &       20      &       290     &       20      &       39      &       5       &       640     &       90      &       i       &       HDE 269050  &       3       \\
Sk --68$^{\circ}$ 155   &       B0.5\,I &       5.51    &       0.2     &       25.4    &       1.4     &       1630    &       100     &       1530    &       0       &       380     &       20      &       25      &       5       &       620     &       100     &       i       &               &       1       \\
Sk --68$^{\circ}$ 140   &       B0.7\,Ib--IabNwk        &       5.64    &       0.1     &       23.5    &       1.7     &       890     &       100     &       1150    &       0       &       230     &       0       &       33      &       4       &       610     &       80      &       i       &       VFTS 696     &       20      \\
Sk --66$^{\circ}$ 35    &       BC1\,Ia &       5.73    &       0.1     &       22      &       1.1     &       490     &       100     &       810     &       50      &       200     &       30      &       37      &       9       &       810     &       280     &       i       &       HDE 268723  &       21      \\
Sk --68$^{\circ}$ 129   &       B1\,I   &       5.25    &       0.2     &       22.2    &       2.2     &       1300    &       100     &       1380    &       40      &       340     &       30      &       23      &       5       &       620     &       130     &       ii      &       W61 27-56   &       1+      \\
Sk --67$^{\circ}$ 2     &       B1\,Ia$^{+}$    &       5.92    &       0.1     &       19.92   &       2.4     &       320     &       100     &       480     &       20      &       120     &       10      &       51      &       8       &       490     &       80      &       ii      &       HDE 270754  &       21      \\
Sk --67$^{\circ}$ 14    &       B1.5\,Ia        &       5.74    &       0.1     &       22.89   &       2.6     &       910     &       100     &       900     &       10      &       220     &       10      &       37      &       5       &       600     &       100     &       i       &       HDE 268685  &       21      \\
Sk --68$^{\circ}$ 26    &       BC2\,Ia &       5.71    &       0.1     &       18.16   &       1.4     &       250     &       140     &       390     &       0       &       100     &       0       &       38      &       4       &       430     &       50      &       i       &               &       24      \\
Sk --69$^{\circ}$ 140   &       B4\,I   &       4.73    &       0.2     &       15      &       1.9     &       890     &       100     &       1130    &       10      &       160     &       10      &       11      &       2       &       540     &       60      &       ii      &               &       1       \\
Sk --70$^{\circ}$ 16    &       B4\,I   &       4.62    &       0.2     &       15      &       2.4     &       910     &       100     &       1010    &       10      &       110     &       10      &       11      &       2       &       580     &       70      &       ii      &               &       1       \\
NGC2004\,ELS\,3 &       B5\,Ia  &       5.1     &       0.2     &       14.45   &       1.6     &       400     &       100     &       440     &       10      &       110     &       10      &       41      &       9       &       1080    &       180     &       ii      &       R 109     &       12      \\
Sk --68$^{\circ}$ 8     &       B5\,Ia$^{+}$    &       5.54    &       0.2     &       14.2    &       2.4     &       270     &       100     &       300     &       0       &       20      &       0       &       40      &       4       &       410     &       110     &       iv      &       HDE 268729  &       1       \\
\hline
\\              
\end{tabular}
\begin{tablenotes}
\item{Reference 1 indicates the stellar parameters come from spectral type calibrations. The addition of the $\text{plus}\text{}$ sign indicates this star was not in the original ULLYSES sample, but has been added from archival spectra. Further numbers indicate the source of the stellar parameters from 2 \citealp{Crowther2002}, 3 \citealp{Evans2004}, 4 \citealp{Walborn2004}, 5 \citealp{Trundle2004}, 6 \citealp{Massey2004}, 7 \citealp{Trundle2005}, 8 \citealp{Dufton2005}, 9 \citealp{Massey2005}, 10 \citealp{Hunter2005}, 11 \citealp{Heap2006}, 12 \citealp{Trundle2007}, 13 \citealp{Hunter2007}, 14 \citealp{Massey2009}, 15 \citealp{RiveroGonzalez2012a}, 16 \citealp{Massey2013}, 17 \citealp{Bouret2013}, 18 \citealp{Hainich2014}, 19 \citealp{Bestenlehner2014}, 20 \citealp{McEvoy2015}, 21 \citealp{Urbaneja2017}, 22 \citealp{RamirezAgudelo2017}, 23 \citealp{Ramachandran2018}, 24 \citealp{Urbaneja2018}, 25 \citealp{Gvaramadze2018}, 26 \citealp{Castro2018}, 27 \citealp{Dufton2019}, 28 \citealp{Mahy2020}, 29 \citealp{Bouret2021}, and 30 \citealp{Pauli2022}}
\end{tablenotes}
\end{table}

\end{landscape}

\begin{landscape}

\begin{table}
\tiny
\caption{Stellar parameters and spectral types from ULLYSES metadata along with terminal wind speeds for giants and bright giants in the LMC.}
\label{tbl:Results-LMC-giants}

\begin{tabular}{lllllllllllllllllll}
\hline
\hline

\thispagestyle{plain}

ID & Sp. Type   & log$(\frac{L}{L_{\odot}})$ & $\Delta$L & $T_{\rm{eff}}$ & $\Delta T_{\rm{eff}}$ & $v_{\infty}$(DM) & $\Delta v$ & $v_{\infty}$(SEI) & $\Delta v$ & $v_{\rm{turb}}$ & $\Delta v$ & $M_{\rm{evol}}$ & $\Delta M_{\rm{evol}}$ & $v_\mathrm{esc}$ & $\Delta v_\mathrm{esc}$ & Q & Alias & Ref  \cr
  &   & [dex] &  & [kK] & & [$\rm{km}\, \rm{s^{-1}}$] & & [$\rm{km}\, \rm{s^{-1}}$] & & [$\rm{km}\, \rm{s^{-1}}$] & & [$M_{\odot}$] & & [$\rm{km}\, \rm{s^{-1}}$] & \\
\hline
Sk --67$^{\circ}$ 211   &       O2\,III(f*)     &       6.34    &       0.1     &       52.5    &       2.8     &       2880    &       40      &       3470    &       50      &       870     &       100     &       118     &       23      &       1650    &       200     &       i       &       HDE 269810  &       4       \\
Sk --66$^{\circ}$ 172   &       O2\,III(f*)+OB  &       6.08    &       0.2     &       49.6    &       5.0     &       3010    &       80      &       3150    &       20      &       190     &       10      &       67      &       19      &       1410    &       250     &       i       &               &       1+      \\
LH 114-7        &       O2\,III(f*)+OB  &       5.65    &       0.2     &       49.6    &       2.4     &       2870    &       100     &       3210    &       20      &       220     &       20      &       54      &       12      &       1440    &       190     &       ii      &               &       1       \\
VFTS 267        &       O3\,III--I(n)f* &       5.96    &       0.1     &       44.1    &       2.2     &       2530    &       100     &       2420    &       60      &       600     &       50      &       62      &       8       &       1270    &       140     &       ii      &               &       22      \\
W61 28-23       &       O3.5\,III(f*)   &       5.65    &       0.1     &       47      &       2.4     &       2970    &       40      &       2840    &       30      &       290     &       30      &       50      &       6       &       1370    &       120     &       i       &               &       15      \\
FMB2009 88      &       O4\,III|(f)     &       5.71    &       0.2     &       42.4    &       2.2     &       2760    &       100     &       2310    &       40      &       580     &       40      &       41      &       8       &       1270    &       200     &       i       &       FMB2009 88      &       1       \\
Sk --67$^{\circ}$ 108   &       O4--5\,III      &       5.94    &       0.2     &       41.4    &       2.7     &       2470    &       10      &       2370    &       40      &       590     &       40      &       49      &       12      &       1150    &       210     &       i       &               &       1       \\
N11 ELS 38      &       O5\,III(f$^{+}$)        &       5.67    &       0.2     &       40.5    &       2.2     &       2500    &       50      &       2290    &       40      &       570     &       40      &       32      &       6       &       1080    &       180     &       i       &       PGMW 3100    &       15      \\
N11 ELS 18      &       O6\,II(f$^{+}$) &       5.76    &       0.2     &       38.2    &       2.1     &       2190    &       170     &       2120    &       40      &       530     &       40      &       38      &       8       &       1020    &       180     &       i       &       PGMW 3053    &       1       \\
VFTS 440        &       O6--6.5\,II(f)  &       5.63    &       0.2     &       33.8    &       3.5     &       2190    &       80      &       1990    &       130     &       500     &       140     &       248     &       90      &       1140    &       260     &       i       &       Mk 47      &       22      \\
Sk --71$^{\circ}$ 19    &       O6\,III &       5.1     &       0.2     &       38.2    &       1.7     &       1180    &       100     &       1840    &       50      &       460     &       90      &       25      &       3       &       1180    &       130     &       iii     &               &       1       \\
Sk --71$^{\circ}$ 50    &       O6.5\,III       &       5.57    &       0.2     &       37.1    &       2.1     &       1980    &       10      &       1810    &       40      &       300     &       40      &       32      &       6       &       1010    &       170     &       i       &               &       1       \\
Sk --68$^{\circ}$ 16    &       O7\,III &       5.68    &       0.2     &       36.1    &       2.0     &       2220    &       100     &       2380    &       20      &       220     &       20      &       34      &       7       &       970     &       170     &       ii      &               &       1       \\
BI 272  &       O7\,II  &       5.47    &       0.2     &       36.1    &       2.0     &       220     &       100     &       290     &       20      &       70      &       20      &       29      &       5       &       1020    &       170     &       iv      &               &       1       \\
LMC X-4 &       O8\,III &       5.02    &       0.2     &       34      &       1.3     &       230     &       100     &       1080    &       30      &       160     &       50      &       21      &       2       &       1110    &       150     &       iii     &               &       1       \\
Sk --67$^{\circ}$ 106   &       O8\,III(f)      &       5.9     &       0.2     &       34      &       1.2     &       1680    &       10      &       1930    &       30      &       390     &       20      &       29      &       6       &       890     &       140     &       ii      &       HDE 269525  &       1       \\
BI 173  &       O8\,II  &       5.6     &       0.1     &       34.5    &       1.2     &       2320    &       100     &       2560    &       80      &       640     &       60      &       36      &       4       &       900     &       80      &       ii      &               &       14      \\
Sk --67$^{\circ}$ 101   &       O8\,II(f)       &       5.62    &       0.2     &       34      &       1.2     &       2250    &       100     &       2350    &       60      &       460     &       50      &       31      &       7       &       890     &       150     &       ii      &               &       1       \\
Sk --70$^{\circ}$ 79    &       B0\,III &       5.63    &       0.2     &       28.7    &       2.4     &       1450    &       0       &       1300    &       40      &       260     &       40      &       30      &       7       &       710     &       130     &       i       &               &       1       \\
LH 9-89 &       B0\,IIIn        &       5.05    &       0.1     &       26.7    &       1.9     &       1240    &       100     &       1260    &       10      &       140     &       10      &       20      &       2       &       740     &       80      &       ii      &       N11 ELS 33  &       15      \\
\hline
\\              
\end{tabular}
\begin{tablenotes}
\item{}
\end{tablenotes}
\end{table}

\begin{table}
\tiny
\caption{Stellar parameters and spectral types from ULLYSES metadata along with terminal wind speeds for dwarfs and sub-giants in the LMC.}
\label{tbl:Results-LMC-dwarfs}

\begin{tabular}{lllllllllllllllllll}
\hline
\hline

ID & Sp. Type   & log$(\frac{L}{L_{\odot}})$ & $\Delta$L & $T_{\rm{eff}}$ & $\Delta T_{\rm{eff}}$ & $v_{\infty}$(DM) & $\Delta v$ & $v_{\infty}$(SEI) & $\Delta v$ & $v_{\rm{turb}}$ & $\Delta v$ & $M_{\rm{evol}}$ & $\Delta M_{\rm{evol}}$ & $v_\mathrm{esc}$ & $\Delta v_\mathrm{esc}$ & Q & Alias & Ref  \cr
  &   & [dex] &  & [kK] & & [$\rm{km}\, \rm{s^{-1}}$] & & [$\rm{km}\, \rm{s^{-1}}$] & & [$\rm{km}\, \rm{s^{-1}}$] & & [$M_{\odot}$] & & [$\rm{km}\, \rm{s^{-1}}$] & \\
\hline
VFTS 72 &       O2\,V--III(n)((f*))     &       5.97    &       0.1     &       54.8    &       2.5     &       2920    &       100     &       3020    &       20      &       320     &       20      &       78      &       16      &       1570    &       190     &       i       &       BI 253     &       15      \\
BI 237  &       O2\,V((f*))     &       5.83    &       0.1     &       53.2    &       2.4     &       2910    &       100     &       3240    &       10      &       240     &       10      &       68      &       13      &       1520    &       170     &       i       &               &       15      \\
N11 ELS 60      &       O3\,V((f*))     &       5.63    &       0.1     &       48      &       2.4     &       2870    &       100     &       3070    &       20      &       180     &       20      &       50      &       7       &       1410    &       120     &       ii      &       PGMW 3058    &       15      \\
W61 28-5        &       O4\,V((f$^{+}$))        &       5.51    &       0.1     &       44      &       2.4     &       2600    &       70      &       2230    &       40      &       530     &       40      &       41      &       5       &       1320    &       110     &       i       &               &       15      \\
PGMW 3120       &       O5.5\,V((f*))   &       5.99    &       0.2     &       40.9    &       2.0     &       2690    &       20      &       2550    &       60      &       490     &       50      &       50      &       13      &       1120    &       210     &       i       &               &       1       \\
Sk --66$^{\circ}$ 19    &       O7\,V   &       5.30    &       0.2     &       37.9    &       3.8     &       1150    &       70      &       1370    &       50      &       340     &       40      &       27      &       5       &       1190    &       200     &       ii      &       N11 ELS 2   &       1+      \\
Sk --67$^{\circ}$ 118   &       O7\,V   &       5.69    &       0.2     &       37.9    &       1.8     &       1950    &       100     &       2360    &       50      &       560     &       50      &       36      &       7       &       1040    &       180     &       iii     &               &       1       \\
N11 ELS 13      &       O8\,V   &       5.74    &       0.2     &       35.9    &       1.6     &       1970    &       100     &       2260    &       30      &       280     &       30      &       37      &       9       &       920     &       160     &       ii      &       BI 42, PGMW 3223   &       1       \\
Sk --67$^{\circ}$ 191   &       O8\,V   &       5.39    &       0.2     &       35.9    &       1.5     &       2040    &       100     &       2340    &       20      &       260     &       20      &       27      &       4       &       990     &       150     &       ii      &               &       1       \\
BI 184  &       O8Ve    &       5.28    &       0.1     &       34      &       1.7     &       200     &       100     &       210     &       0       &       50      &       0       &       27      &       2       &       940     &       90      &       iv      &       NGC206 FS 119  &       23      \\
VFTS 66 &       O9\,V+B0.2\,V   &       4.54    &       0.1     &       32.8    &       1.5     &       260     &       100     &       280     &       0       &       70      &       10      &       17      &       1       &       1070    &       70      &       iv      &               &       28      \\
HV 5622 &       B0\,V   &       4.59    &       0.2     &       30.5    &       2.8     &       1090    &       100     &       1180    &       10      &       70      &       10      &       15      &       2       &       1020    &       140     &       iii     &               &       1       \\
NGC206 FS 170   &       B1\,IV  &       4.55    &       0.1     &       24.0    &       1.5     &       240     &       100     &       300     &       0       &       70      &       0       &       13      &       1       &       700     &       70      &       iv      &               &       23      \\
\hline
\\              
\end{tabular}
\begin{tablenotes}
\item{}
\end{tablenotes}
\end{table}

\end{landscape}

\begin{landscape}

\thispagestyle{plain}

\begin{table}
\tiny
\centering
\caption{Stellar parameters and spectral types from ULLYSES metadata along with terminal wind speeds for supergiants in the SMC.}
\label{tbl:Results-SMC-sgiants}

\begin{tabular}{lllllllllllllllllll}
\hline
\hline

ID & Sp. Type   & log$(\frac{L}{L_{\odot}})$ & $\Delta$L & $T_{\rm{eff}}$ & $\Delta T_{\rm{eff}}$ & $v_{\infty}$(DM) & $\Delta v$ & $v_{\infty}$(SEI) & $\Delta v$ & $v_{\rm{turb}}$ & $\Delta v$ & $M_{\rm{evol}}$ & $\Delta M_{\rm{evol}}$ & $v_\mathrm{esc}$ & $\Delta v_\mathrm{esc}$ & Q & Alias & Ref  \cr
  &   & [dex] &  & [kK] & & [$\rm{km}\, \rm{s^{-1}}$] & & [$\rm{km}\, \rm{s^{-1}}$] & & [$\rm{km}\, \rm{s^{-1}}$] & & [$M_{\odot}$] & & [$\rm{km}\, \rm{s^{-1}}$] & \\
\hline
AzV 26  &       O6\,I(f)        &       6.14    &       0.2     &       38.0    &       3.6     &       1540    &       100     &       2150    &       40      &       480     &       40      &       41      &       8       &       1000    &       180     &       ii      &       Sk 18      &       16+     \\
AzV 15  &       O6.5\,I(f)      &       5.83    &       0.1     &       39.0    &       2.0     &       2340    &       70      &       2340    &       10      &       140     &       10      &       48      &       5       &       1050    &       90      &       i       &       Sk 10      &       29      \\
AzV 220 &       O6.5f?p &       5.20    &       0.2     &       37.5    &       3.5     &       1540    &       100     &       1930    &       60      &       480     &       140     &       18      &       3       &       1100    &       150     &       iii     &       NGC346 1019    &       11      \\
AzV 232 &       O7\,Iaf$^{+}$   &       5.89    &       0.1     &       33.5    &       1.7     &       1550    &       30      &       1310    &       30      &       310     &       30      &       48      &       5       &       880     &       80      &       i       &       Sk 80      &       29      \\
AzV 83  &       O7\,Iaf$^{+}$   &       5.54    &       0.1     &       32.8    &       1.6     &       1190    &       50      &       1030    &       20      &       250     &       20      &       33      &       4       &       860     &       80      &       i       &               &       29      \\
2dFS 163        &       O8\,Ib(f)       &       4.82    &       0.2     &       32.6    &       3.3     &       1190    &       100     &       1170    &       10      &       180     &       10      &       17      &       3       &       1080    &       170     &       ii      &               &       1       \\
AzV 479 &       O9\,Ib  &       5.82    &       0.2     &       29.0    &       3.1     &       1400    &       100     &       1620    &       20      &       430     &       20      &       35      &       8       &       800     &       160     &       i       &       Sk 155     &       1       \\
AzV 372 &       O9.5\,Iabw      &       5.62    &       0.1     &       28.0    &       1.4     &       1480    &       190     &       1510    &       20      &       280     &       10      &       35      &       4       &       710     &       70      &       i       &       Sk 116     &       3       \\
AzV 287 &       O9.5\,I &       6.16    &       0.2     &       29.8    &       3.0     &       1540    &       70      &       1620    &       30      &       420     &       30      &       52      &       7       &       820     &       130     &       i       &       Sk 101     &       1+      \\
AzV 456 &       O9.5\,Ib        &       5.81    &       0.1     &       29.5    &       1.5     &       1660    &       100     &       1610    &       0       &       140     &       10      &       44      &       5       &       760     &       80      &       ii      &       Sk 143     &       3       \\
AzV 70  &       O9.5\,Iw        &       5.68    &       0.1     &       28.5    &       1.4     &       2010    &       100     &       2010    &       20      &       180     &       20      &       37      &       4       &       740     &       80      &       ii      &       Sk 35      &       3       \\
AzV 317 &       B0\,Iw  &       5.40    &       0.2     &       26.9    &       2.9     &       1540    &       50      &       1380    &       20      &       250     &       20      &       22      &       4       &       820     &       170     &       i       &       Sk 107     &       26      \\
AzV 215 &       B0\,Ib  &       5.63    &       0.1     &       27.0    &       1.4     &       1540    &       100     &       1540    &       40      &       230     &       30      &       35      &       4       &       680     &       70      &       ii      &       Sk 76      &       5       \\
AzV 235 &       B0\,Iaw &       5.72    &       0.1     &       27.5    &       1.4     &       1520    &       100     &       1430    &       50      &       440     &       50      &       37      &       5       &       690     &       70      &       ii      &       Sk 82      &       3       \\
AzV 16  &       B0[e]   &       5.75    &       0.2     &       27.2    &       2.7     &       360     &       10      &       380     &       50      &       150     &       30      &       32      &       7       &       700     &       140     &       iv      &       Sk 11, R4  &       1       \\
AzV 488 &       B0.5\,Iaw       &       5.74    &       0.1     &       27.5    &       1.4     &       1250    &       100     &       1200    &       20      &       250     &       20      &       39      &       5       &       700     &       70      &       i       &       Sk 159     &       3       \\
AzV 104 &       B0.5\,Ia        &       5.31    &       0.1     &       27.5    &       1.4     &       310    &       100     &       600     &       40      &       170     &       30      &       24      &       2       &       700     &       70      &       iii     &               &       7       \\
AzV 242 &       B0.7\,Iaw       &       5.67    &       0.1     &       25.0    &       1.3     &       1120    &       100     &       1120    &       20      &       200     &       10      &       35      &       4       &       620     &       60      &       ii      &       Sk 85      &       7       \\
AzV 266 &       B1\,I   &       5.09    &       0.1     &       18.2    &       0.9     &       1330    &       100     &       1340    &       10      &       60      &       10      &       16      &       1       &       570     &       40      &       iv      &       Sk 95      &       26      \\
Sk 191  &       B1.5\,Ia        &       5.77    &       0.1     &       22.5    &       1.1     &       630     &       10      &       620     &       10      &       160     &       20      &       39      &       4       &       550     &       60      &       i       &               &       5       \\
AzV 210 &       B1.5\,Ia        &       5.41    &       0.1     &       20.5    &       1.0     &       810     &       60      &       860     &       30      &       140     &       20      &       26      &       3       &       580     &       60      &       ii      &       Sk 73      &       5       \\
AzV 78  &       B1.5\,Ia$^{+}$  &       5.92    &       0.2     &       21.5    &       2.1     &       440    &       100     &       590     &       10      &       150     &       20      &       48      &       6       &       530     &       80      &       ii      &       Sk 40, R9, HD5045  &       5+      \\
AzV 18  &       B2\,Ia  &       5.44    &       0.1     &       19.0    &       1.0     &       430     &       100     &       460     &       20      &       180     &       30      &       42      &       4       &       480     &       70      &       iv      &       Sk 13      &       5       \\
AzV 393 &       B2\,Ia  &       5.80    &       0.2     &       19.0    &       1.9     &       320     &       100     &       330     &       10      &       80      &       10      &       45      &       5       &       500     &       80      &       iv      &       Sk 124     &       1       \\
AzV 472 &       B2\,Ia  &       5.31    &       0.2     &       19.0    &       2.1     &       150    &       100     &       200     &       20      &       80     &       30      &       18      &       4       &       600     &       120     &       iv      &       Sk 150     &       1, 8    \\
NGC330 ROB B22  &       B2\,IIe &       4.72    &       0.2     &       21.2    &       1.9     &       150       &       100     &       120       &       10      &       50       &       10       &       10      &       1       &       630     &       80      &       iv      &               &       1+      \\
NGC330 ELS 4    &       B2.5\,Ib        &       4.77    &       0.1     &       17.0    &       0.9     &       540     &       100     &       300     &       60      &       120     &       50      &       12      &       1       &       570     &       20      &       iv      &               &       12      \\
AzV 362 &       B3\,Ia  &       5.50    &       0.1     &       14.0    &       0.7     &       290     &       100     &       290     &       10      &       120     &       30      &       49      &       1       &       330     &       30      &       iv      &       Sk 114     &       5       \\
NGC330 ELS 2    &       B3\,Ib  &       4.73    &       0.1     &       15.0    &       0.7     &       550     &       100     &       580     &       10       &       20      &       10       &       -       &       -       &       -       &       -       &       iv      &               &       5       \\
AzV 234 &       B3\,Iab &       4.91    &       0.1     &       15.7    &       0.8     &       290     &       100     &       280     &       10      &       110     &       20      &       -       &       -       &       -       &       -       &       iv      &       Sk 81      &       27      \\
AzV 324 &       B4\,Iab &       4.89    &       0.2     &       14.6    &       1.5     &       310     &       100     &       310     &       10       &       120     &       30      &       -       &       -       &       -       &       -       &       iv      &               &       1       \\
AzV 22  &       B5\,Ia  &       5.04    &       0.1     &       14.5    &       0.7     &       300     &       10      &       300     &       20      &       120     &       20      &       -       &       -       &       -       &       -       &       iv      &               &       5       \\
\hline
\\              
\end{tabular}
\begin{tablenotes}
\item{}
\end{tablenotes}
\end{table}

\end{landscape}

\begin{landscape}

\thispagestyle{plain}

\begin{table}
\tiny
\centering
\caption{Stellar parameters and spectral types from ULLYSES metadata along with terminal wind speeds for giants and bright giants in the SMC.}
\label{tbl:Results-SMC-giants}

\begin{tabular}{lllllllllllllllllll}
\hline
\hline

ID & Sp. Type   & log$(\frac{L}{L_{\odot}})$ & $\Delta$L & $T_{\rm{eff}}$ & $\Delta T_{\rm{eff}}$ & $v_{\infty}$(DM) & $\Delta v$ & $v_{\infty}$(SEI) & $\Delta v$ & $v_{\rm{turb}}$ & $\Delta v$ & $M_{\rm{evol}}$ & $\Delta M_{\rm{evol}}$ & $v_\mathrm{esc}$ & $\Delta v_\mathrm{esc}$ & Q & Alias & Ref  \cr
  &   & [dex] &  & [kK] & & [$\rm{km}\, \rm{s^{-1}}$] & & [$\rm{km}\, \rm{s^{-1}}$] & & [$\rm{km}\, \rm{s^{-1}}$] & & [$M_{\odot}$] & & [$\rm{km}\, \rm{s^{-1}}$] & \\
\hline
NGC346 MPG 355  &       O2\,III(f*)     &       6.04    &       0.1     &       51.7    &       2.6     &       3120    &       100     &       3120    &       30      &       200     &       20      &       57      &       3       &       1520    &       140     &       ii      &               &       17      \\
AzV 80  &       O4--6(n)fp      &       5.80    &       0.1     &       38.0    &       1.9     &       1700    &       100     &       1810    &       20      &       120     &       10      &       47      &       5       &       1030    &       100     &       i       &               &       11      \\
AzV 75  &       O5\,IIIf$^{+}$  &       5.94    &       0.1     &       38.5    &       1.9     &       2160    &       50      &       2150    &       10      &       860     &       140     &       55      &       4       &       1070    &       80      &       i       &       Sk 38      &       29      \\
AzV 95  &       O7\,III((f))    &       5.46    &       0.1     &       38.0    &       1.9     &       1950    &       100     &       1940    &       30      &       780     &       140     &       33      &       3       &       1040    &       100     &       i       &               &       29      \\
AzV 207 &       O7\,III((f))    &       5.34    &       0.1     &       37.0    &       1.9     &       1440    &       100     &       1470    &       20      &       590     &       120     &       29      &       3       &       1020    &       90      &       iii     &               &       6       \\
AzV 440 &       O7.5\,III       &       5.28    &       0.1     &       37.0    &       1.9     &       1240    &       100     &       1280    &       10      &       510     &       120     &       28      &       2       &       1030    &       90      &       iii     &               &       9       \\
AzV 69  &       OC7.5\,III((f)) &       5.61    &       0.1     &       33.9    &       1.7     &       1870    &       20      &       1990    &       10      &       260     &       30      &       36      &       4       &       890     &       90      &       i       &       Sk 34      &       29      \\
AzV 47  &       O8\,III((f))    &       5.44    &       0.1     &       35.0    &       1.8     &       2210    &       100     &       2220    &       20      &       110     &       10       &       31      &       3       &       940     &       90      &       iii     &               &       29      \\
AzV 454 &       O8.5\,III       &       5.56    &       0.2     &       32.9    &       3.3     &       1550    &       100     &       490     &       30      &       200     &       30      &       24      &       4       &       900     &       170     &       ii      &       Sk 142     &       1+      \\
AzV 321 &       O9\,IInp        &       5.28    &       0.2     &       31.8    &       3.2     &       950     &       30      &       920     &       40      &       20      &       10       &       23      &       4       &       870     &       160     &       i       &               &       1       \\
AzV 307 &       O9\,III &       5.15    &       0.1     &       30.0    &       1.5     &       150    &       100     &       300    &       20      &       120     &       20      &       22      &       2       &       810     &       70      &       iii     &               &       29      \\
AzV 6   &       O9\,III &       5.81    &       0.2     &       31.8    &       3.2     &       150    &       10     &       550    &       50      &       220     &       20      &       34      &       8       &       850     &       180     &       iii     &               &       1       \\
AzV 327 &       O9.5\,II-Ibw    &       5.54    &       0.1     &       30.0    &       1.5     &       1600    &       100     &       1580    &       50      &       570     &       50      &       32      &       4       &       770     &       70      &       ii      &               &       29      \\
AzV 423 &       O9.5\,II(n)     &       5.34    &       0.1     &       28.2    &       1.4     &       1660    &       100     &       1660    &       10      &       130     &       10      &       26      &       3       &       730     &       70      &       ii      &       Sk 132     &       26      \\
AzV 170 &       O9.7\,III       &       5.14    &       0.2     &       30.5    &       2.9     &       170    &       100     &       470     &       80      &       320     &       190     &       15      &       2       &       930     &       170     &       iv      &               &       29      \\
Sk 173  &       B0.7\,IIe       &       5.00    &       0.1     &       24.0    &       1.2     &       150     &       100     &       170     &       20      &       70     &       30      &       17      &       2       &       640     &       50      &       iv      &               &       23      \\
AzV 224 &       B1\,III &       4.92    &       0.2     &       24.0    &       2.4     &       260     &       100     &       270     &       20      &       20      &       10       &       15      &       2       &       670     &       100     &       iv      &               &       1       \\
AzV 175 &       B1\,IIw &       5.24    &       0.2     &       24.0    &       2.4     &       300     &       100     &       330     &       70      &       130     &       40      &       17      &       4       &       600     &       110     &       iv      &       Sk 64      &       1       \\
AzV 216 &       B1\,III &       5.00    &       0.2     &       26.0    &       2.9     &       140       &       10      &       150       &       10      &       60       &       10       &       16      &       3       &       850     &       160     &       iv      &               &       5       \\
\hline
\\              
\end{tabular}
\begin{tablenotes}
\item{}
\end{tablenotes}
\end{table}

\end{landscape}

\begin{landscape}

\thispagestyle{plain}

\begin{table}
\tiny
\centering
\caption{Stellar parameters and spectral types from ULLYSES metadata along with terminal wind speeds for dwarfs and sub-giants in the SMC. The full ID of OGLE* is OGLE-J004942.75-731717.7. The full ID of MOA* is MOA-J010321.3-720538.}
\label{tbl:Results-SMC-dwarfs}

\begin{tabular}{lllllllllllllllllll}
\hline
\hline

ID & Sp. Type   & log$(\frac{L}{L_{\odot}})$ & $\Delta$L & $T_{\rm{eff}}$ & $\Delta T_{\rm{eff}}$ & $v_{\infty}$(DM) & $\Delta v$ & $v_{\infty}$(SEI) & $\Delta v$ & $v_{\rm{turb}}$ & $\Delta v$ & $M_{\rm{evol}}$ & $\Delta M_{\rm{evol}}$ & $v_\mathrm{esc}$ & $\Delta v_\mathrm{esc}$ & Q & Alias & Ref  \cr
  &   & [dex] &  & [kK] & & [$\rm{km}\, \rm{s^{-1}}$] & & [$\rm{km}\, \rm{s^{-1}}$] & & [$\rm{km}\, \rm{s^{-1}}$] & & [$M_{\odot}$] & & [$\rm{km}\, \rm{s^{-1}}$] & \\
\hline
AzV 476 &       O2--3\,V+OB     &       5.71    &       0.1     &       39.8    &       2.0     &       2620    &       100     &       2620    &       0       &       140     &       0       &       43      &       5       &       1070    &       100     &       i       &               &       26      \\
AzV 388 &       O4\,V   &       5.54    &       0.1     &       43.1    &       2.2     &       2160    &       100     &       2180    &       30      &       130     &       20      &       39      &       4       &       1250    &       120     &       ii      &               &       17      \\
AzV 177 &       O4\,V((f))      &       5.17    &       0.2     &       43.9    &       4.4     &       1540    &       100     &       2670    &       10      &       80      &       10      &       28      &       5       &       1240    &       160     &       ii      &               &       17      \\
NGC346 ELS 7    &       O4\,V((f))      &       5.51    &       0.1     &       42.1    &       2.1     &       2550    &       100     &       2560    &       90      &       340     &       60      &       37      &       4       &       1190    &       110     &       iii     &       NGC346\,MPG\,324        &       17      \\
OGLE*   &       O4--7\,V((f))e  &       5.46    &       0.2     &       40.9    &       4.0     &       1540    &       140     &       1370    &       30      &       340     &       20      &       27      &       5       &       1190    &       170     &       i       &               &       1+      \\
AzV 480 &       O4--7\,Ve       &       5.80    &       0.2     &       40.9    &       4.4     &       160    &       100     &       200     &       30      &       80      &       10      &       36      &       7       &       1300    &       190     &       iv      &               &       1+      \\
AzV 377 &       O5\,V((f))      &       5.54    &       0.1     &       45.5    &       2.3     &       150    &       10     &       140     &       10      &       30      &       10      &       42      &       5       &       1370    &       120     &       iv      &               &       6       \\
NGC346\,MPG\,342        &       O5--6\,V(f)     &       5.63    &       0.2     &       39.9    &       4.2     &       1540    &       100     &       1660    &       160     &       410     &       90      &       39      &       8       &       1280    &       220     &       iii     &               &       1+      \\
NGC346\,MPG\,368        &       O5.5\,V((f$^{+}$))      &       5.38    &       0.1     &       39.3    &       2.0     &       2240    &       100     &       2210    &       20      &       130     &       20      &       31      &       3       &       1100    &       100     &       ii      &               &       17      \\
AzV 243 &       O6\,V   &       5.59    &       0.1     &       39.6    &       2.0     &       2110    &       100     &       2110    &       10      &       150     &       10      &       38      &       4       &       1080    &       100     &       ii      &       Sk 84      &       17      \\
2DFS-5066       &       O6\,V((f))      &       5.52    &       0.2     &       39.9    &       4.0     &       1540    &       100     &       1940    &       10      &       80      &       10      &       29      &       5       &       1220    &       190     &       iii     &               &       1+      \\
MOA*    &       O6\,V+O4--5III  &       5.40    &       0.2     &       41.0    &       4.0     &       1540    &       100     &       1880    &       10      &       70      &       10      &       34      &       7       &       1220    &       220     &       ii      &       [MWD2000]\,H53-47       &       1+      \\
NGC346 ELS 28   &       OC6\,Vz &       5.15    &       0.1     &       39.6    &       2.0     &       230     &       100     &       240     &       10      &       60      &       10      &       27      &       2       &       1210    &       100     &       iv      &       NGC346\,MPG\,113        &       17      \\
AzV 446 &       O6.5\,V &       5.25    &       0.1     &       39.7    &       2.0     &       330     &       100     &       310     &       0       &       40      &       10      &       29      &       3       &       1170    &       110     &       iv      &               &       17      \\
NGC346\,MPG\,602        &       O6.5\,V((f))    &       5.18    &       0.2     &       38.9    &       3.8     &       280       &       20     &       180       &       30       &       70       &       20       &       22      &       4       &       1190    &       160     &       iv      &       NGC346 1026    &       1+      \\
NGC346 ELS 51   &       O7\,Vz  &       4.80    &       0.1     &       38.7    &       1.9     &       230     &       100     &       280     &       10      &       70      &       0       &       22      &       2       &       1200    &       80      &       iv      &       NGC346\,MPG\,523        &       17      \\
NGC346\,MPG\,396        &       O7\,V   &       5.22    &       0.2     &       37.9    &       3.8     &       270       &       10     &       180       &       20       &       70       &       20       &       23      &       4       &       1140    &       150     &       iv      &               &       1+      \\
NGC346\,MPG\,487        &       O8\,V   &       5.12    &       0.2     &       38.5    &       3.6     &       160       &       10     &       280       &       20       &       70       &       20       &       22      &       4       &       1180    &       210     &       iv      &               &       17+     \\
NGC346 ELS 50   &       O8\,Vn  &       4.64    &       0.1     &       36.3    &       1.8     &       150    &       10     &       150     &       10      &       80      &       30      &       19      &       2       &       1150    &       70      &       iv      &       NGC346 MPG 299 &       17      \\
AzV 267 &       O8\,V   &       4.90    &       0.1     &       35.7    &       1.8     &       1300    &       100     &       1300    &       20      &       110     &       10      &       21      &       2       &       1080    &       90      &       iii     &               &       17      \\
NGC346 ELS 22   &       O9\,V   &       4.89    &       0.2     &       34.8    &       3.4     &       260       &       10     &       160       &       10       &       60       &       10       &       16      &       2       &       1070    &       130     &       iv      &       NGC346 MPG 682 &       17+     \\
AzV 326 &       B0\,IV  &       4.81    &       0.1     &       32.4    &       1.6     &       470     &       100     &       360     &       20      &       80      &       20      &       18      &       1       &       970     &       90      &       iii     &               &       17      \\
AzV 189 &       B0\,IV  &       4.81    &       0.1     &       32.3    &       1.6     &       360     &       100     &       340     &       10      &       80      &       20      &       18      &       1       &       960     &       90      &       iv      &               &       17      \\
NGC346 ELS 26   &       B0\,IV (Nstr)   &       4.93    &       0.1     &       31.0    &       1.6     &       230     &       120     &       360     &       20      &       90      &       40      &       19      &       1       &       870     &       80      &       iv      &       NGC346 MPG 12  &       17      \\
NGC346 ELS 43   &       B0\,V   &       4.71    &       0.1     &       33.0    &       1.7     &       230     &       100     &       260     &       10      &       60      &       10      &       17      &       1       &       1020    &       90      &       iv      &       NGC346 MPG 11  &       13      \\
AzV 43  &       B0.2\,V &       5.13    &       0.1     &       28.5    &       1.4     &       1300    &       100     &       1320    &       10      &       70      &       10      &       21      &       2       &       760     &       70      &       ii      &               &       17      \\
AzV 304 &       B0.5\,V &       4.86    &       0.2     &       27.5    &       2.8     &       330     &       100     &       300     &       10      &       70      &       20      &       14      &       2       &       880     &       160     &       iv      &               &       1, 10      \\
NGC330 ROB A1   &       B0.5\,Ve        &       4.86    &       0.2     &       28.6    &       2.9     &       270       &       20     &       160       &       20       &       20       &       10       &       14      &       2       &       990     &       160     &       iv      &               &       1+      \\
2DFS-3694       &       B1\,IV  &       4.69    &       0.1     &       24.0    &       1.2     &       360     &       100     &       330     &       20      &       10      &       0       &       14      &       1       &       660     &       50      &       iv      &               &       23      \\
NGC346 ELS 35   &       B1\,V   &       4.69    &       0.2     &       27.3    &       2.7     &       320     &       100     &       300     &       10      &       40      &       10      &       13      &       2       &       910     &       160     &       iv      &               &       1       \\
\hline
\\              
\end{tabular}
\begin{tablenotes}
\item{}
\end{tablenotes}
\end{table}

\end{landscape}

\section{Trends with lower-quality measurements} \label{sec: low qual}

\begin{figure}[t!]
    \centering
    \includegraphics[scale=0.2]{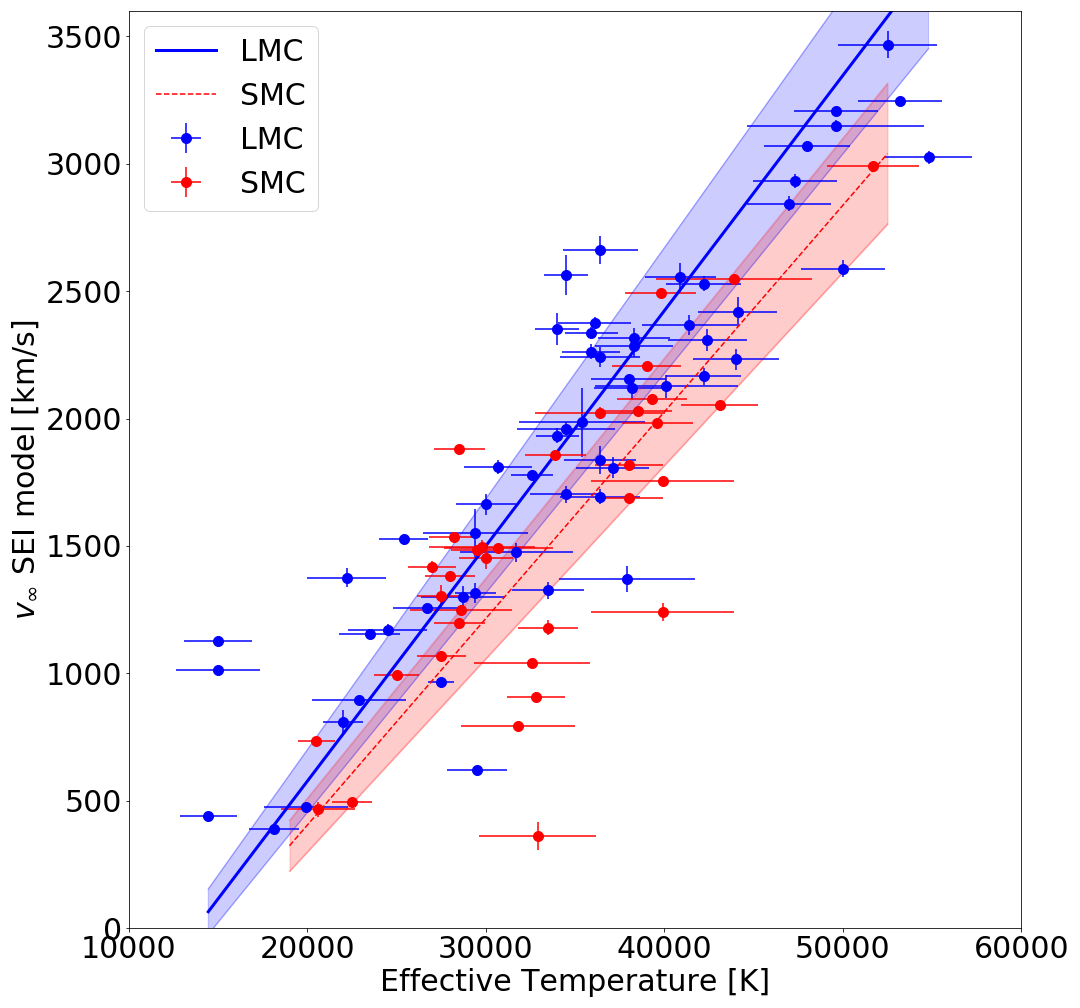}
    \caption{As Figure \ref{fig: Vinf-Teff-SEI} but including $v_{\infty}$ measurements up to quality rank ii.} 
    \label{fig: Vinf-Teff-appblue}
\end{figure}

\begin{figure}[t!]
    \centering
    \includegraphics[scale=0.2]{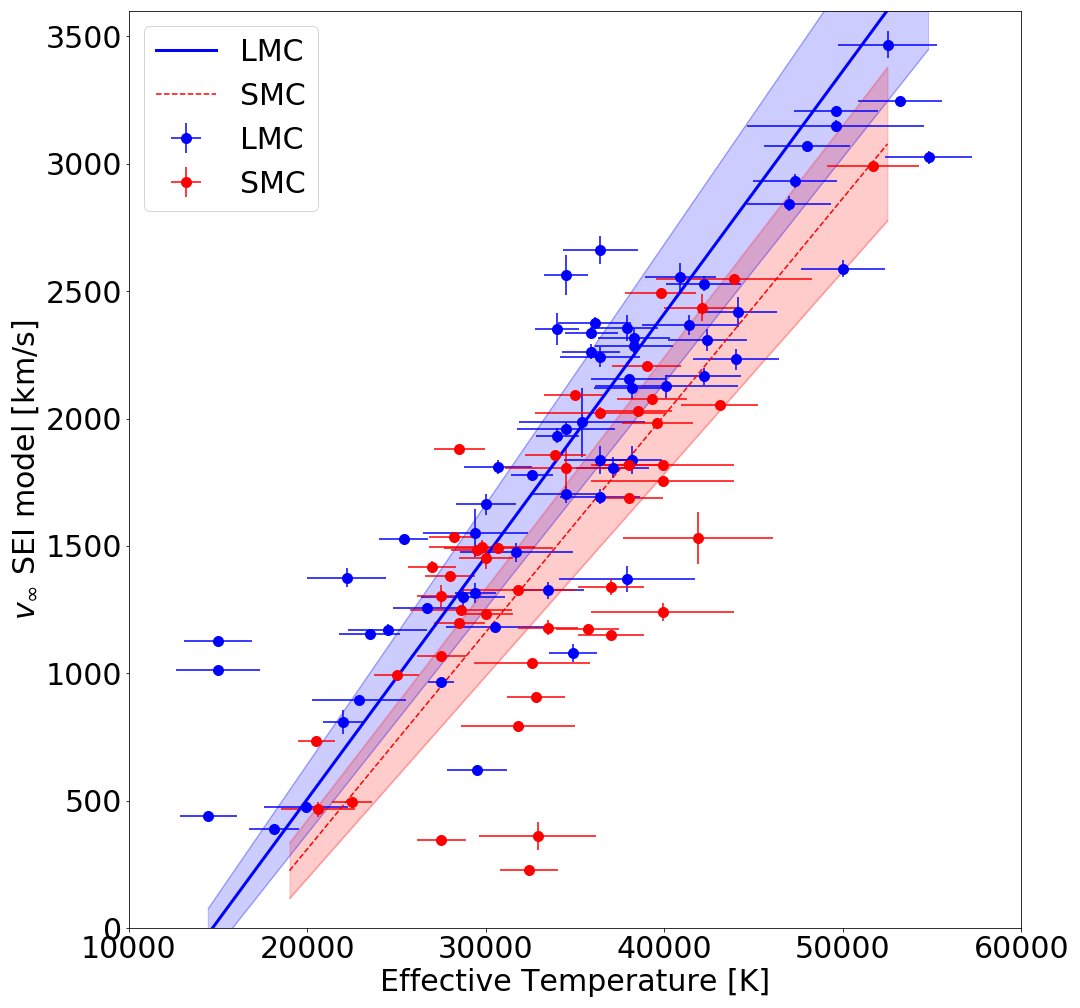}
    \caption{As Figure \ref{fig: Vinf-Teff-SEI}, but including $v_{\infty}$ measurements up to quality rank iii.} 
    \label{fig: Vinf-Teff-appyell}
\end{figure}

\begin{figure}[t!]
    \centering
    \includegraphics[scale=0.2]{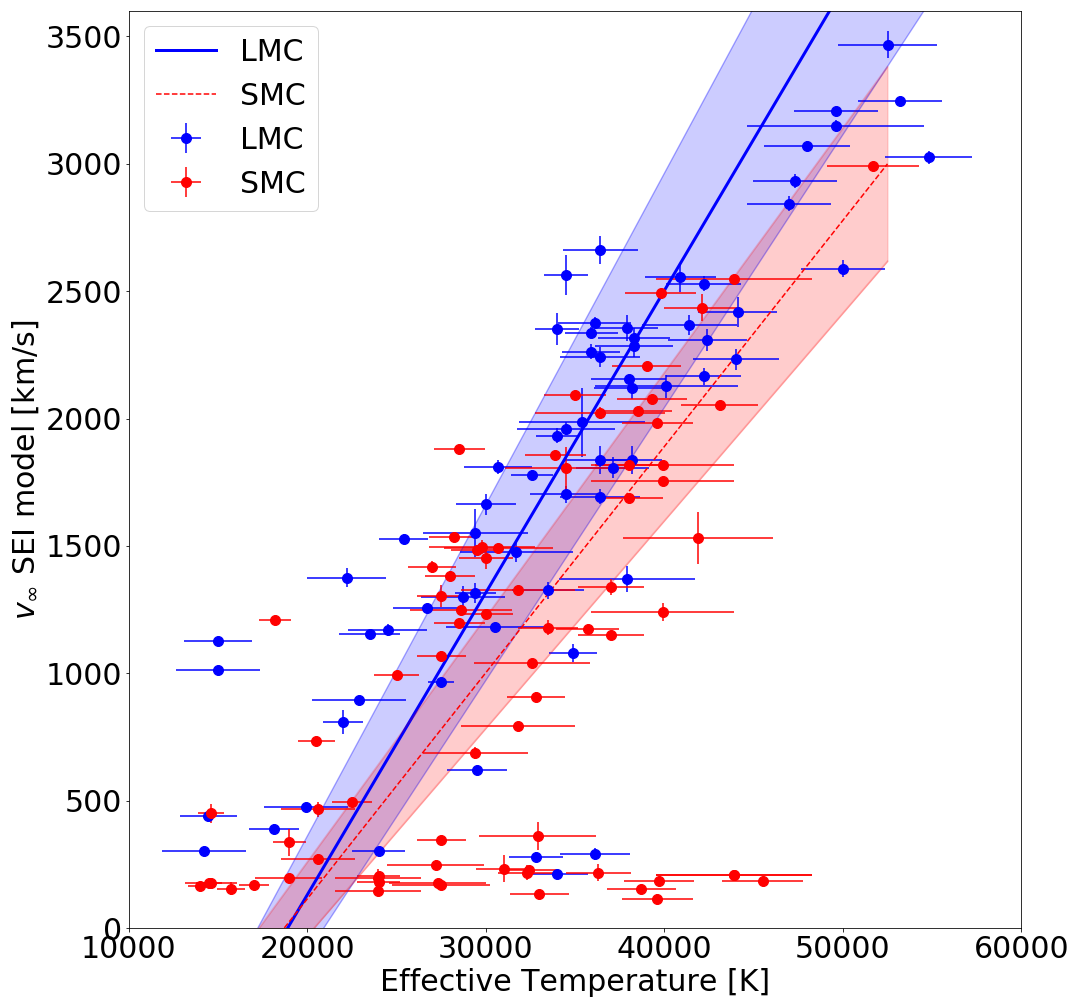}
    \caption{As Figure \ref{fig: Vinf-Teff-SEI}, but including $v_{\infty}$ measurements up to quality rank iv.} 
    \label{fig: Vinf-Teff-appyred}
\end{figure}

Here we present the same analysis as in Sect. \ref{sec: vinf-teff}, but now including measurements of poorer quality ranks. The inclusion of each consecutive decreasing rank is shown in Figs. \ref{fig: Vinf-Teff-appblue} to \ref{fig: Vinf-Teff-appyred}. Including rank ii measurements increases the slopes in both environments, but both agree with rank i findings within the errors. This might indicate that including wind speed measurements from unsaturated profiles, which are lower than the true $v_{\infty}$ , has a systematic effect on the slope. This effect is exacerbated when adding rank iii measurements, which may be even farther from the true $v_{\infty}$. In a few cases, the $v_{\infty}$ lies far above the slope. The spectral types collected from the literature for these stars may not be correct. Updating spectral types for the ULLYSES sample is beyond the scope of this study. Rank iv measurements are for stars with little to no wind signatures, a large portion of which can be seen in Fig. \ref{fig: Vinf-Teff-appyred} to have high temperatures. This is likely linked to the weak-wind problem where we are unable to diagnose $v_{\infty}$ from UV resonance lines. Therefore, there is little motivation to include these measurements when investigating $v_{\infty}$. In Sect. \ref{sec: vinf-teff} we only used quality rank i measurements, that is, saturated line profiles.

\section{Trends by luminosity class} \label{lum-class}

\begin{figure}[t!]
    \centering
    \includegraphics[scale=0.2]{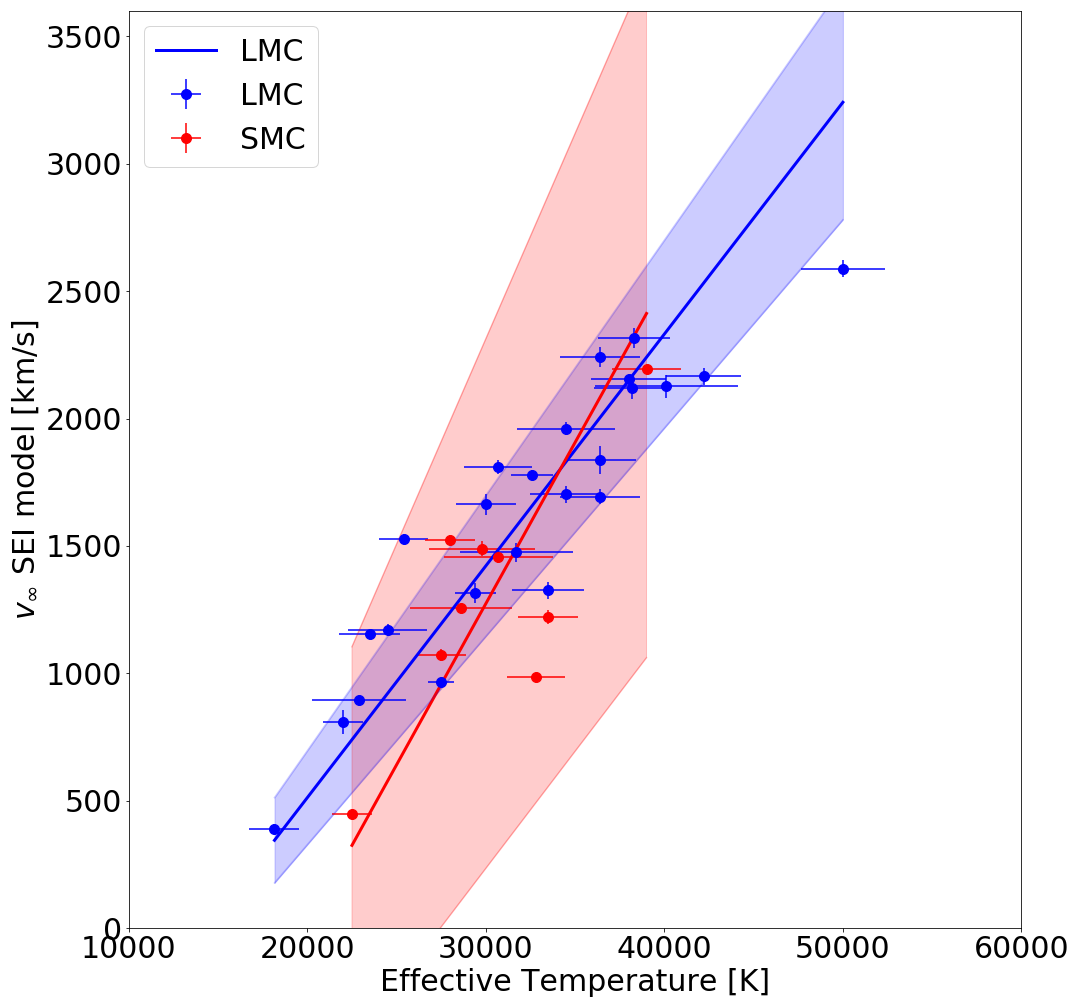}
    \caption{As Figure \ref{fig: Vinf-Teff-SEI}, but showing only $v_{\infty}$ measurements for supergiant stars.} 
    \label{fig: Vinf-Teff-appsgiants}
\end{figure}

\begin{figure}[t!]
    \centering
    \includegraphics[scale=0.2]{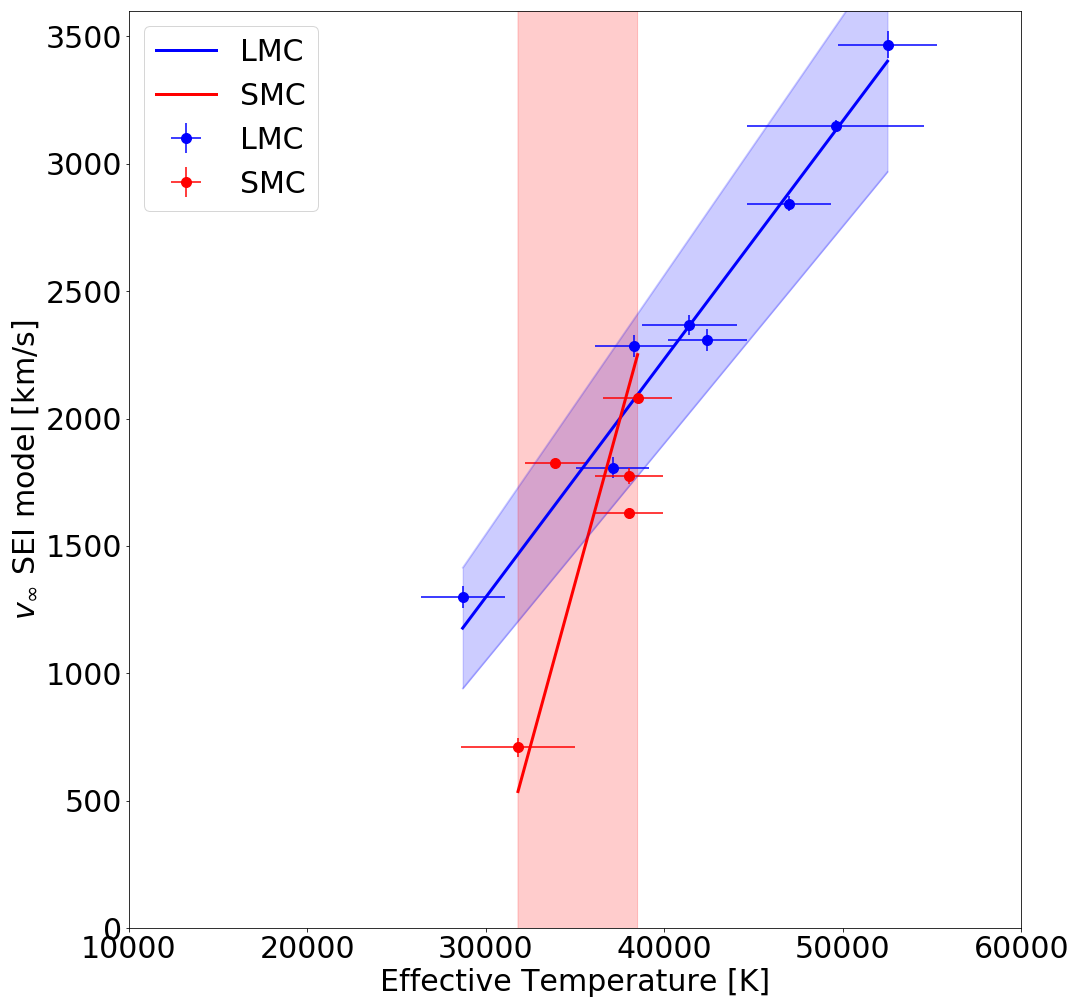}
    \caption{As Figure \ref{fig: Vinf-Teff-SEI}, but showing only $v_{\infty}$ measurements for giant stars.} 
    \label{fig: Vinf-Teff-appgiants}
\end{figure}

\begin{figure}[t!]
    \centering
    \includegraphics[scale=0.2]{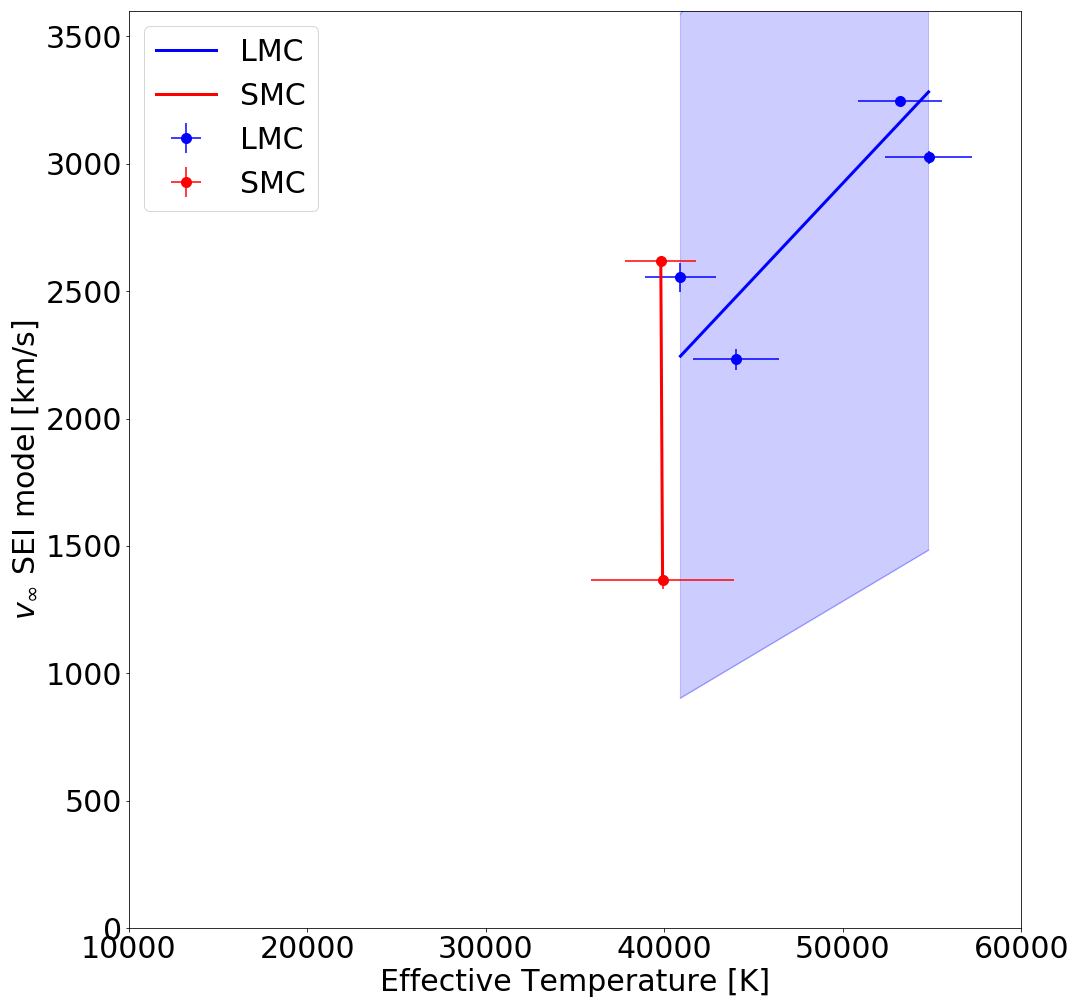}
    \caption{As Figure \ref{fig: Vinf-Teff-SEI}, but showing only $v_{\infty}$ measurements for dwarf stars.} 
    \label{fig: Vinf-Teff-appdwarfs}
\end{figure}

Here we show the effect of analysing the sample split by luminosity class. The sub-samples for supergiants alone (luminosity class I) in the LMC and SMC are shown in Fig. \ref{fig: Vinf-Teff-appsgiants}. The sub-samples for giants and bright giants alone (luminosity class III) in the LMC and SMC are shown in Fig. \ref{fig: Vinf-Teff-appgiants}. The sub-samples for dwarfs and sub-dwarfs alone (luminosity class V) in the LMC and SMC are shown in Fig. \ref{fig: Vinf-Teff-appdwarfs}. For the LMC, the slopes are not affected sufficiently to motivate analysing the trends by luminosity class. The low number of high-quality measurements prevent this comparison in the SMC.

\section{Galactic samples} \label{App-Galaxy}

\begin{table}
\centering
\caption{ODR fit coefficients to relations between $v_{\infty}$ and $T_{\rm{eff}}$ as described in Eq. 3. }
\label{tbl:Results-App}
\vspace{2mm}
\begin{tabular}{lcccc}
\hline
\hline

Region & a & b & RMS & Sample  \cr
  & [$10^{-2}$] & [$\rm{km}\, \rm{s^{-1}}$] & [$\rm{km}\, \rm{s^{-1}}$] &  \\
\hline

LMC     &       $9.2 \pm 0.6$   &       $1270 \pm 200$  &       385      &  LMC-ii  \\
SMC     &       $9.9 \pm 1.1$   &       $1630 \pm 350$  &       452      &  SMC-ii  \\
LMC     &       $9.5 \pm 0.7$   &       $1400 \pm 230$  &       407      &  LMC-iii  \\
SMC     &       $12.6 \pm 1.8$  &       $2600 \pm 560$  &       594      &  SMC-iii  \\
LMC     &       $11.8 \pm 1.1$  &       $2240 \pm 380$  &       645      &  LMC-iv  \\
SMC     &       $11.7 \pm 1.6$  &       $2190 \pm 420$  &       987      &  SMC-iv  \\
\hline
LMC     &       $9.1 \pm 0.9$   &       $1310 \pm 280$  &       268      &  LMC-I  \\
SMC     &       $12.6 \pm 3.5$  &       $2520 \pm 1010$ &       346      &  SMC-I  \\
LMC     &       $9.3 \pm 0.8$   &       $1500 \pm 340$  &       118      &  LMC-III  \\
SMC     &       $25.6 \pm 21.6$ &       $7600 \pm 7920$ &       443      &  SMC-III  \\
LMC     &       $7.5 \pm 3.3$   &       $810 \pm 160$   &       238      &  LMC-V  \\
SMC     &       n/a     &       n/a     &       n/a      &  SMC-V  \\
\hline
GAL     &       $9.8 \pm 0.3$   &       $1230 \pm 960$  &       326      &  G  \\
GAL     &       $10.2 \pm 0.4$  &       $1340 \pm 120$  &       291      &  H  \\
GAL     &       $10.2 \pm 0.3$  &       $1300 \pm 100$  &       278      &  M  \\
\hline
\\              
\end{tabular}
\begin{tablenotes}
\item{The column 'sample' indicates the metallicity environment and whether the slope is measured for a poorer quality rank (lower case roman numerals) or limited to a specific luminosity class (upper case roman numerals).}
\end{tablenotes}
\end{table}

\begin{figure}[t!]
    \centering
    \includegraphics[scale=0.2]{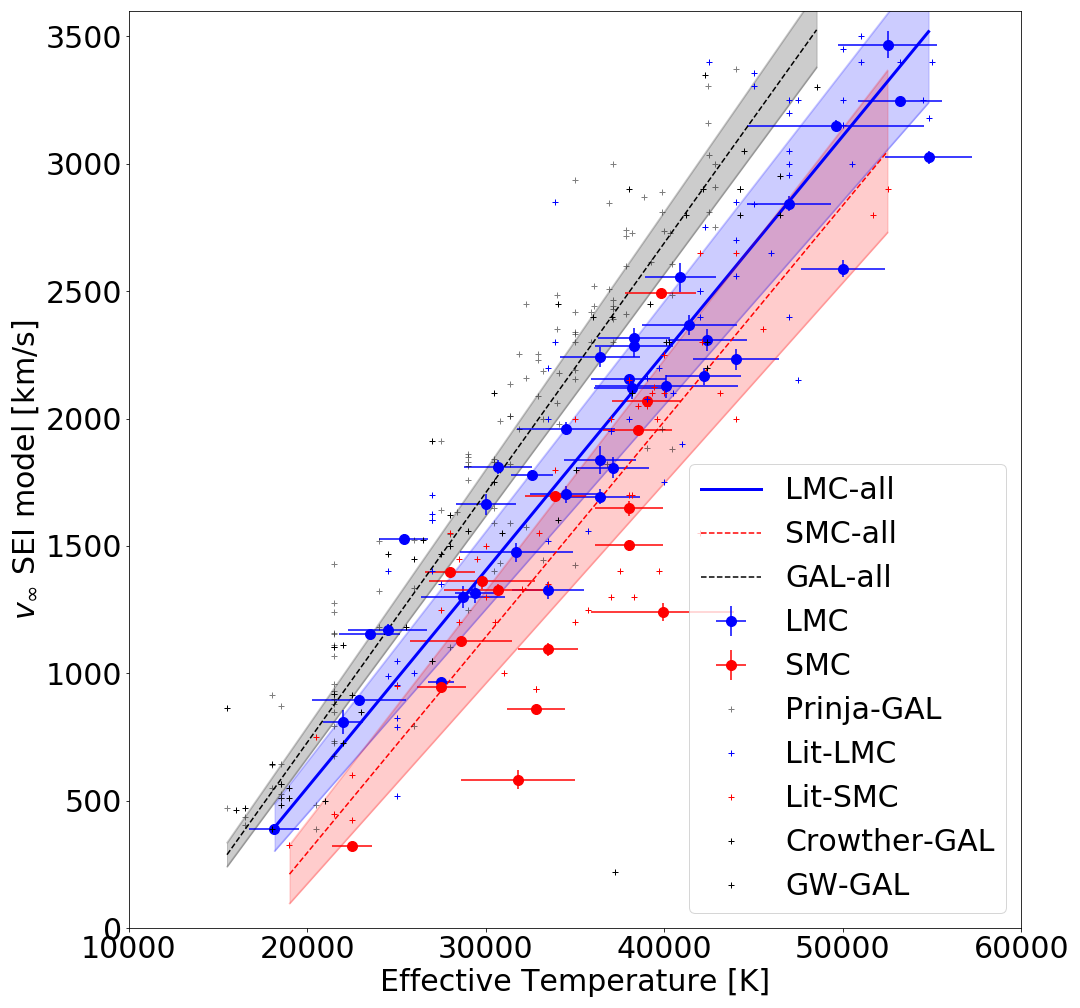}
    \caption{As Figure \ref{fig: Vinf-Teff-Zdep}, but also including Galactic samples from \cite{Groenewegen1989} and \cite{Crowther2006}.} 
    \label{fig: Vinf-Teff-moreGAL}
\end{figure}

We present the inclusion of the \cite{Groenewegen1989} and \cite{Crowther2006} samples in the Galactic relation. This is sample G in Table \ref{tbl:Results-App}. The slope of the relation including these samples agrees slightly better with our findings in the LMC and SMC, but the quality of the $v_{\infty}$ measurements in these studies is difficult to verify.

\begin{figure}[t!]
    \centering
    \includegraphics[scale=0.25]{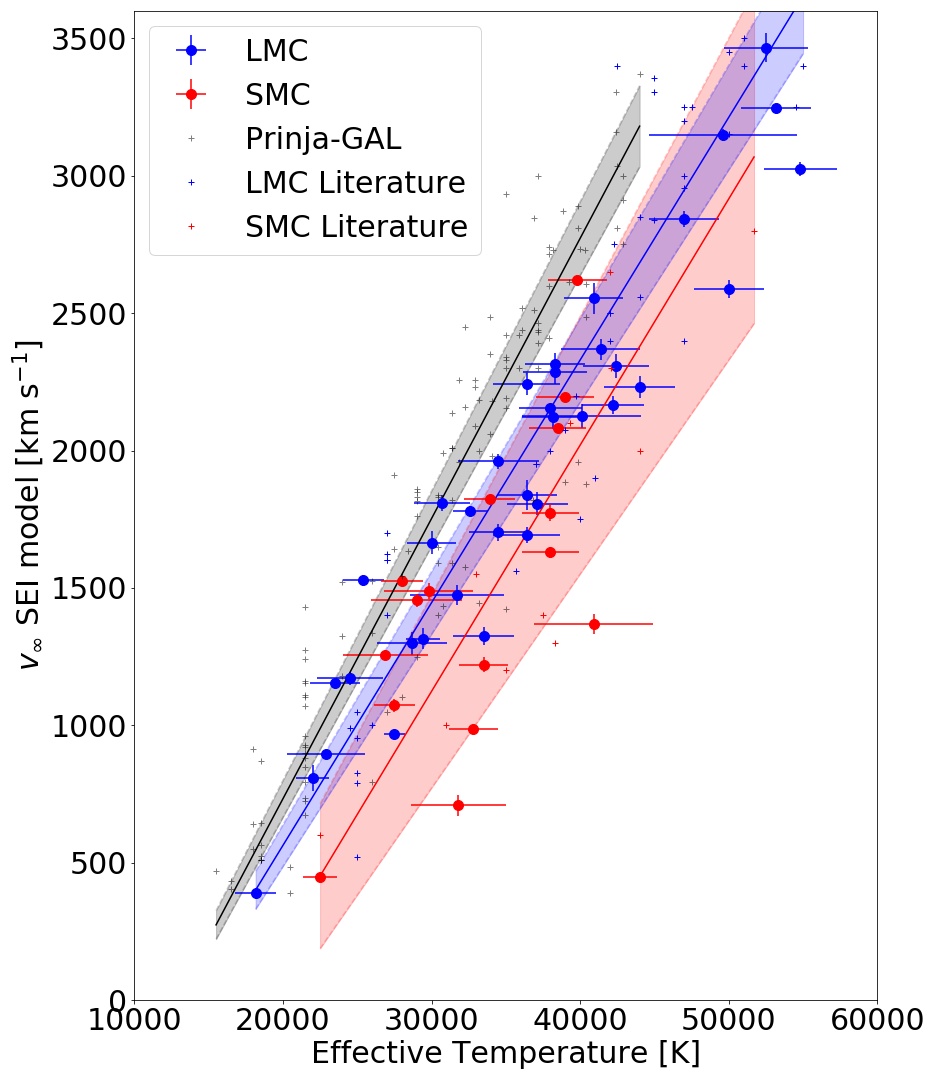}
    \caption{As Figure \ref{fig: Vinf-Teff-Zdep}, but using the full Galatic O star sample with saturated wind profiles from \cite{Prinja1990}.} 
    \label{fig: Vinf-Teff-Holgado}
\end{figure}

As discussed in Sect. \ref{sec: vinf-z}, we updated the stellar parameters of the Galactic O star sample with results obtained through quantitative spectroscopy. The work of \cite{Holgado2020} provides best-fit parameters from quantitative spectroscopy of optical spectra using FASTWIND models. The sample of \cite{Holgado2020}  comprises more than 400 stars from the Galactic O star catalogue and includes 93\% of the \cite{Prinja1990} sample with saturated \ion{C}{IV} $\lambda\lambda$1548-1550 profiles. Only 72\% of the stars have spectroscopic best-fit solutions, however, because 20 objects were identified as double-line spectroscopic binaries (SB2). For the purpose of updating the stellar parameters for this study, we used effective temperatures from \cite{Holgado2020} when available and excluded SB2 stars from the sample. This is presented as sample H in Table \ref{tbl:Results-App}. As there is no UV component to the \cite{Holgado2020} study, we used the terminal wind speeds found \cite{Prinja1990}. We also tested the effect of using the full \cite{Prinja1990} O star sample with saturated wind profiles. In this case, we used the full \cite{Prinja1990} sample by updating spectral types using the Galactic O Star Spectroscopic Survey (GOSSS; \citealp{Sota2011, Sota2014, Apellaniz2016}) and used them to determine stellar parameters from the observational temperature calibrations of \cite{Martins2005}. This is sample M in Table \ref{tbl:Results-App}. The result of this is an overall slight upward revision of the terminal wind speed for a given effective temperature, as shown in Fig. \ref{fig: Vinf-Teff-Holgado}, resulting in a slightly steeper metallicity dependence of $v_{\infty} \propto Z^{0.24 \pm 0.03}$, which is not significantly different from the relation found using the effective temperatures from \cite{Holgado2020}.

\end{appendix}

\end{document}